\documentclass[useAMS,usenatbib]{mn2e}

\usepackage{graphicx}
\usepackage{pdflscape}
\usepackage{subfigure}

\newcommand{\Hb}{H$\beta$}   
\newcommand{\Fe}{$\langle$Fe$\rangle$}  
\newcommand{\Mgb}{Mg$b$}  
\newcommand{\Mgd}{Mg$_2$}   
\newcommand{\MgFe}{[MgFe]}     

\newcommand{\kms}{$\mathrm{km\;s^{-1}}$} 
\newcommand{\kmsmpc}{$\mathrm{km\;s^{-1}\;Mpc^{-1}}$} 
 
\newcommand{\msun}{M$_{\odot}$} 
\newcommand{\mlsun}{$\rm M_{\odot}/L_{\odot}$} 

\newcommand{\age}{$\tau$}
\newcommand{\ZH}{[$Z$/H]}
\newcommand{\aFe}{[$\alpha$/Fe]}

\newcommand{\mlkroupa}{\Upsilon_\mathrm{\ast,Kroupa}} 
 
\newcommand{\mldyn}{\Upsilon_\mathrm{\ast,dyn}} 
\newcommand{\ratml}{\mldyn/\mlkroupa}

\newcommand{\rhohalo}{\rho_\mathrm{DM,dyn}} 
\newcommand{\rhodmkroupa}{\rho_\mathrm{DM,Kroupa}} 
\newcommand{\rhoast}{\rho_\ast} 
\newcommand{\avgrhohalo}{\langle \rhohalo \rangle}
\newcommand{\avgrhodmkroupa}{\langle \rhodmkroupa \rangle}
 
\newcommand{\halofrac}{f_\mathrm{DM,dyn}} 
\newcommand{\dmkroupafrac}{f_\mathrm{DM,Kroupa}}

\newcommand{\reff}{r_\mathrm{eff}} 
\newcommand{\renc}{r_\mathrm{enc}} 
 
\newcommand{\sigeff}{\sigma_\mathrm{eff}}
\newcommand{\pa}{\mathrm{P.A.}}  

\newcommand{\zdm}{z_\mathrm{DM}}

\def\aap{A\&A}
\def\aaps{A\&AS}
\def\aj{AJ}
\def\apj{ApJ}
\def\apjl{ApJ}
\def\apjs{ApJS}

\def\mnras{MNRAS}
\def\pasp{PASP}

\title[Dark matter haloes of early-type galaxies]{The density of dark
  matter haloes of early-type galaxies in low-density environments}

\author[E. M. Corsini et
  al.]{E. M. Corsini$^{1,2}$\thanks{E-mail:
    enricomaria.corsini@unipd.it}, G. A. Wegner$^3$\thanks{Visiting
    Astronomer, MDM Observatory, Kitt Peak, Arizona, operated by a
    consortium of Dartmouth College, University of Michigan,
    Columbia University, The Ohio State University, and Ohio
    University.}, J. Thomas$^4$, R. P. Saglia$^4$ and R. Bender$^{4,5}$\\
$^1$Dipartimento di Fisica e Astronomia `G. Galilei', Universit\`a di Padova, 
  vicolo dell'Osservatorio 3, 35122 Padova, Italy\\ 
$^2$INAF--Osservatorio Astronomico di Padova, 
  vicolo dell'Osservatorio 2, 35122 Padova, Italy\\ 
$^3$Department of Physics and Astronomy, 6127 Wilder Laboratory,   
  Dartmouth College, Hanover, NH 03755-3528, USA\\  
$^4$Max-Planck-Institut f\"ur extraterrestrische Physik,  
  Giessenbachstra{\ss}e, D-85748 Garching, Germany\\  
$^5$Universit\"ats-Sternwarte M\"unchen, Scheinerstra{\ss}e~1,  
  D-81679 M\"unchen, Germany}  

\begin{document}

\date{Accepted 2016 November 9. Received 2016 November 9; in original form 2016 July 20}

\pagerange{\pageref{firstpage}--\pageref{lastpage}} \pubyear{2002}

\maketitle

\label{firstpage}

\begin{abstract}
  New photometric and long-slit spectroscopic observations are
  presented for NGC~7113, PGC~1852, and PGC~67207 which are three
  bright galaxies residing in low-density environments. The
  surface-brightness distribution is analysed from the $K_S$-band images
  taken with adaptive optics at the Gemini North Telescope and
  the $ugriz$-band images from the Sloan Digital Sky Survey while the
  line-of-sight stellar velocity distribution and line-strength Lick
  indices inside the effective radius are measured along several
  position angles. The age, metallicity, and $\alpha$-element
  abundance of the galaxies are estimated from single
  stellar-population models. In spite of the available morphological
  classification, images show that PGC~1852 is a barred spiral which
  we do not further consider for mass modelling. The structural
  parameters of the two early-type galaxies NGC~7113 and PGC~67207 are
  obtained from a two-dimensional photometric decomposition and the
  mass-to-light ratio of all the (luminous and dark) mass that follows
  the light is derived from orbit-based axisymmetric dynamical
  modelling together with the mass density of the dark matter
  halo. The dynamically derived mass that follows the light is about a
  factor of 2 larger than the stellar mass derived using
  stellar-population models with Kroupa initial mass function. Both
  galaxies have a lower content of halo dark matter with respect to
  early-type galaxies in high-density environments and in agreement
  with the predictions of semi-analytical models of galaxy formation.
\end{abstract}


\section{Introduction}
\label{sec_Introduction}

A comprehensive picture of galaxy formation and evolution requires
understanding the assembly and growth of massive early-type galaxies
(ETGs). These systems dominate the bright end of the luminosity
function \citep{Bernardi2010}, contain most of the present-day stellar
mass \citep{Fukugita1998}, and can be observed out to high redshifts
\citep{Daddi2005}.
The central dark matter (DM) density, stellar mass-to-light ratio, and
distribution of stellar orbits provide valuable clues on the epoch
and mechanism of the assembly of ETGs.

Both theoretical arguments and numerical simulations predict a
relationship between the concentration and assembly epoch of the DM
haloes with denser haloes formed earlier \citep{Wechsler2002}. The
halo mass distribution also depends on the interplay between baryons
and DM. Indeed the central DM and stellar mass profile could become
similar as the result of violent relaxation \citep{Hilz2012}, whereas
adiabatic contraction is able to increase the amount of DM in the
galaxy centres lowering the amount of the stellar mass
\citep{Gnedin2004}.
The stellar mass-to-light ratio is set by the initial mass function
(IMF) which determines the mass scale and chemical enrichment of
galaxies too. The systematic variation of the IMF with the galaxy mass
in ETGs results from the analysis of integrated spectra
\citep{vanDokkum2010, Spiniello2012, Ferreras2013, LaBarbera2013,
  Smith2015b}, stellar dynamics \citep{Thomas2011, Cappellari2012,
  Wegner2012, Barnabe2013, Tortora2013}, scaling relations
\citep{Dutton2013}, and gravitational lensing \citep{Auger2010,
  Treu2010} and it is interpreted with an enhanced fraction of
low-mass stars in more massive galaxies. Recently,
\citet{MartinNavarro2015} have found that the IMF in massive ETGs
strongly depends on galactocentric distance with the excess of
low-mass stars confined into the central regions and implying a
difference between the formation process of the core and outer
regions.

Numerical simulations and semi-analytical models predict, on the other
hand, that environment also plays an important role in setting the
properties of galaxies \citep[e.g.,][]{Kauffmann2004}. DM haloes in
low-density environments assemble and virialize later with respect to
their equal mass counterparts in high-density environments
\citep[e.g.,][]{Hahn2007}. As a consequence, galaxies in low-density
regions could have younger stellar populations than their mass
analogous in clusters of galaxies \citep[e.g.,][]{Bernardi1998,
  Delucia2007}.
In general, the population of galaxies in voids is dominated by small
and faint late-type spirals which are gas rich and star forming. But,
some massive ETGs with stellar mass $M_\ast>10^{10}$ \msun\ and no
active star formation in the last several Gyrs are also present
\citep{Croton2005, Beygu2016}. They have similar properties on average
of galaxies living in field or clusters and hosted by DM haloes of
similar mass suggesting comparable mass assembly and quenching process
\citep{Croton2008, Penny2015}. In particular, the lack of difference
in the star formation rates of massive galaxies independently of their
environment points to quenching being a mass-related effect
\citep{FraserMcKelvie2016}. Studying in detail the mass distribution
and properties of stellar populations of galaxies in voids provides a
direct insight on their assembly and quenching histories.
However, except for a few objects \citep[e.g.,][]{Rieder2013}, not
much is known about the density of DM haloes of galaxies in voids and
low-density environments. In particular, orbit-based dynamical models
aimed at constraining the dark and luminous mass distribution within
the half-light radius $\reff$ of large samples of ETGs from long-slit
or integral-field stellar kinematics considered only galaxies in
field, groups, and clusters (\citealt{Cappellari2006, Thomas2007,
  Thomas2014}, Zhu et al., in preparation). Here, we present our first
attempt to tackle this issue.

This study complements our earlier works on ETGs in Coma and Abell~262
clusters \citep{Thomas2007, Thomas2011, Wegner2012}, which we selected
to be respectively a rich and a poor nearby cluster at about the same
distance. As a consequence, the spatial resolution of the photometric
and kinematic data, which consist of broad-band images and long-slit
spectra obtained along various position angles, is comparable for all
the sample galaxies \citep{Mehlert2000, Wegner2002, Wegner2012,
  Corsini2008}. We derived the dynamical mass-to-light ratio of the
mass that follows light and DM halo parameters from axisymmetric
orbit-based dynamical models in combination with the stellar
mass-to-light ratio, age, metallicity, and $\alpha$-elements abundance
from single stellar population models. The DM fraction of total mass
inside $\reff$ is significant ($\halofrac \approx 0.2-0.5$) in most of
the sample ETGs. Their haloes are about 10 times denser than in
spirals of the same stellar mass and assembled at larger redshifts
($\zdm\approx1-3$) and match with constraints from stellar
populations. The stars of some galaxies appear to be younger than the
halo, which indicates a secondary star-formation episode after the
main halo assembly.  The dynamical mass that follows the light is
larger than expected for a Kroupa IMF, especially in galaxies with
high velocity dispersion $\sigeff$ inside $\reff$. Some of them show a
negligible DM fraction. This could indicate a Salpeter IMF in massive
ETGs. Alternatively, some of the DM in massive galaxies could follow
the light very closely suggesting a significant degeneracy between
dark and luminous matter or the assumption of a constant stellar
mass-to-light ratio inside a galaxy should be relaxed.  These findings
can be easily compared to those for ETGs in low-density environments
since the data sets are similar and we analysed them with the same
dynamical and stellar population models.

The structure of the paper is as follows. In Section~\ref{sec_sample},
we discuss the selection of the sample galaxies. In
Section~\ref{sec_imaging}, we describe the imaging data set and
perform the isophotal analysis and photometric decomposition. In
Section~\ref{sec_spectroscopy}, we present the spectroscopic data set
and measure the stellar kinematics and line-strength indices along
different axes and derive the properties of the stellar
populations. In Section~\ref{sec_dynamics}, we discuss the results of
the dynamical modelling. In Section~\ref{sec_conclusions}, we draw our
conclusions.

\section{Sample selection}
\label{sec_sample}

As a pilot project, we have endeavoured to study in more detail three
objects randomly chosen from the list of \citet{Wegner2008} who
focused on 26 ETGs within and bordering three prominent nearby voids
at $cz\approx5000-10000$ \kms\ in the Second Center for Astrophysics
Redshift Survey \citep[CfA2;][]{Geller1989}. The galaxies were
selected to be ellipticals or lenticulars based on their morphology in
the blue images of the Second Palomar Observatory Sky Survey (POSS-II)
and to be located in underdense regions, where the smoothed galaxy
number density is lower than the mean density for the CfA2 survey as
defined in \citet{Grogin1999}. \citet{Wegner2008} found that while the
oldest ETGs in voids and Coma cluster have comparable ages, the age
range of void ETGs is wider and several objects are younger and show
undergoing star formation.

Table \ref{tab_galaxies} lists the properties of the selected
galaxies. We quote the $B$- and $R$-band total magnitudes that are about a
magnitude brighter than the values given in \citet{Wegner2008}; this
stems mainly from our choice of distances, that are based on systemic
velocities with respect to the cosmic microwave background reference
frame and the concordance cosmology ($H_0=73$ \kmsmpc,
$\Omega_{\rm m}=0.27$, $\Omega_\Lambda=0.73$).
We also give the smoothed density contrast derived by
\citet{Grogin1999}, who included PGC~1852 into their lowest density
($n/\overline{n}<0.5$) and NGC~7113 and PGC~67207 into their higher
density ($0.5<n/\overline{n}<1$) void subsamples. Different void
algorithms qualitatively give similar results but differ a lot in the
details of the edge locations \citep{Colberg2008}. According to
\citet{Kreckel2011}, only a few of the galaxies from \citet{Grogin1999}
are located in extremely underdense void interiors.
\citet{Kreckel2011} could derive the filtered density contrast only
for the galaxies falling in sky regions with an extended uniform
coverage of the Sloan Digital Sky Survey (SDSS) Data Release 7
\citep[DR7;][]{Abazajian2009}. This is not the case for our objects
and we decided to estimate their filtered density contrast
($-0.4<\delta<1.6$) from an outlier-resistant linear regression
correlating the density contrast of \citet{Grogin1999} to that of
\citet{Kreckel2011}. We conclude that our sample galaxies are more
probably reside in the low-density outskirts of voids rather than
being located in their desolately underdense central regions.

PGC~1852 was initially included based on earlier POSS morphological
classification as an elliptical galaxy, but improved imaging reveals
that it is a barred spiral and it has been excluded from the dynamical
analysis, although we include the photometric and spectroscopic data
here.

\renewcommand{\tabcolsep}{3pt}
\begin{table*}
\caption{Properties of the sample galaxies.\label{tab_galaxies}}
\begin{center}
\begin{small}
\begin{tabular}{llccccccc}
\hline 
\noalign{\smallskip}   
\multicolumn{1}{c}{Object} & 
\multicolumn{1}{c}{Alt. name} &
\multicolumn{1}{c}{Type} &  
\multicolumn{1}{c}{$V_{\rm CMB}$} & 
\multicolumn{1}{c}{$D_A$} &   
\multicolumn{1}{c}{$(m-M)_L$} &   
\multicolumn{1}{c}{$M_B$} &
\multicolumn{1}{c}{$M_R$} & 
\multicolumn{1}{c}{$n/\overline{n}$}\\
\multicolumn{1}{c}{} & 
\multicolumn{1}{c}{} &
\multicolumn{1}{c}{} &  
\multicolumn{1}{c}{(\kms)} & 
\multicolumn{1}{c}{(Mpc)} &   
\multicolumn{1}{c}{(mag)} &   
\multicolumn{1}{c}{(mag)} &
\multicolumn{1}{c}{(mag)} &
\multicolumn{1}{c}{} \\   
\multicolumn{1}{c}{(1)} & 
\multicolumn{1}{c}{(2)} &  
\multicolumn{1}{c}{(3)} & 
\multicolumn{1}{c}{(4)} &
\multicolumn{1}{c}{(5)} &
\multicolumn{1}{c}{(6)} & 
\multicolumn{1}{c}{(7)} &
\multicolumn{1}{c}{(8)} &
\multicolumn{1}{c}{(9)} \\   
\noalign{\smallskip}   
\hline
\noalign{\smallskip}       
NGC~7113  & CGCG~$2140.0+1221$ & S0 & 5405 & 72.5 & 34.38 & $-20.07$ & $-21.60$ & 0.68\\
PGC~1852  & CGCG~$0027.9+0536$ & E2 & 6737 & 89.9 & 34.86 & $-19.78$ & $-21.18$ & 0.37\\
PGC~67207 & CGCG~$2140.0+1340$ & E2 & 5044 & 67.7 & 34.23 & $-19.52$ & $-20.97$ & 0.82\\
\noalign{\smallskip}   
\hline
\end{tabular}
\end{small}
\end{center}     
\begin{minipage}{17cm}
{\em Note.\/} Column (1): galaxy name. Column (2): alternative
name. Column (3): morphological type from \citet{Wegner2008}. In spite
of the E2 classification, PGC~1852 is a barred spiral galaxy. Column
(4): systemic velocity with respect to the cosmic microwave background
(CMB) reference frame from the NASA/IPAC Extragalactic Database (NED).
Column (5): angular distance from NED adopting $H_0=73$ \kmsmpc,
$\Omega_{\rm m}=0.27$, and $\Omega_\Lambda=0.73$. Column (6): distance
modulus from the luminosity distance in NED.  Column (7): absolute
magnitude in the Johnson-Cousins $B$ band from the apparent magnitude
within the $\mu_B=26$ mag arcsec$^{-2}$ isophote measured by
\citet{Grogin1999} and after applying the Galactic absorption
correction \citep{Schlafly2011} and $K_B$ correction
\citep{Chilingarian2010} available in NED. Column (8): absolute
magnitude in the Johnson-Cousins $R$ band from the apparent magnitude
within the $\mu_R=26$ mag arcsec$^{-2}$ isophote measured by
\citet{Grogin1999} and after applying the Galactic absorption
correction \citep{Schlafly2011} and $K_R$ correction
\citep{Chilingarian2010} available in NED. Column (9): smoothed
density contrast calculated with an uncertainty $\la0.1$ by
\citet{Grogin1999}.
\end{minipage}
\end{table*}

\section{Broad-band imaging}
\label{sec_imaging}

\subsection{Sloan Digital Sky Survey imaging}
\label{sec_sdss}

We retrieved the $ugriz$-band images of the sample galaxies from the
Data Archive Server (DAS) of the SDSS DR9 \citep{Ahn2012}. 

All the archive images were already bias subtracted, flat-field
corrected, sky subtracted, and flux calibrated according to the
associated calibration information stored in the DAS.
We trimmed the images selecting a field of view (FOV) of
$400\times400$ pixel (corresponding to $2.6\times2.6$ arcmin$^2$)
centred on the galaxies. 

To estimate the goodness of the SDSS sky subtraction, we fitted
elliptical isophotes with the algorithm by \citet{Bender1987} to
measure the radial profile of the azimuthally-averaged surface
brightness of the galaxies at large radii. We masked foreground stars,
nearby and background galaxies, residual cosmic rays, and bad pixels
before fitting the isophotes. As a first step, we allowed to vary the
centres, ellipticities, and position angles of ellipses. Then, we
adopted the centre of the inner ellipses ($a<5$ arcsec) and the
ellipticity and position angle of the outer ones ($a>10$ arcsec). We
assumed the constant value of the surface brightness at large radii
($a\sim70$ arcsec) as the residual sky level to be subtracted from the
image. We measured the standard deviation of the image background
after sky subtraction, $\sigma_{\rm sky}$, in regions free of sources
at the edges of the FOV.

\subsection{Gemini North Telescope imaging}
\label{sec_gemini}

To map the central surface brightness distribution with a higher
spatial resolution with respect to that of the SDSS images, NGC~7113
and PGC~67207 were observed with the Near Infrared Imager
\citep[NIRI;][]{Hodapp2003} in combination with the Altitude Conjugate
Adaptive Optics for the Infrared \citep[ALTAIR;][]{Christou2010} and
laser guide star \citep[LGS;][]{Boccas2006} system on the 8.2 m Gemini
North Telescope on 2011 August 16 and October 16, respectively.

The $f/32$ focus was employed with a $1024\times1024$ pixel ALLADIN-II
InSb array as the detector, which gave a $22.5 \times 22.5 $
arcsec$^2$ FOV with a plate scale of 0.022 arcsec pixel$^{-1}$. A
$K_s$ filter ranging from 1.99 and 2.30 $\mu$m and centred on 2.15
$\mu$m was used for all the observations, the field lens of the
instrument was placed `in', and the Cassegrain rotator was set to
`following'.
NGC~7113 and a PSF standard star at a distance of 1.9 arcmin were
observed with 10 s exposures. A total of 62, 25, and 14 useable
exposures were collected for the galaxy, sky, and PSF standard star,
respectively.
PGC~67207 and a PSF standard star at a distance of 2.7 arcmin were
observed with 15 s and 10 s exposures, respectively. A total of 14,
12, and 12 exposures were secured for the galaxy, sky, and PSF
standard star, respectively.
A five-pointing dither pattern with 2.0 arcsec offsets was used for
each set of observations. The sky frames were obtained at a distance
of about 2.0 arcmin from the target galaxy.
The flux standard star FS~152 was also observed, but its images were
not analysed because the flux calibration was performed by fitting the
surface-brightness radial profiles measured in the Gemini images
to the SDSS ones. Flat-field and dark frames were provided on the same
day as the target observations using the same readout mode and
exposure times.

We reduced the images using the {\sc niri} tasks of the Gemini
{\sc iraf}\footnote{The Imaging Reduction and Analysis Facility ({\sc
    iraf}) is distributed by the National Optical Astronomy
  Observatory, which is operated by the Association of Universities
  for Research in Astronomy (AURA), Inc., under cooperative agreement
  with the National Science Foundation.}  reduction package. Briefly,
these included {\sc nprepare} which set up the image headers, {\sc
  niflat} and {\sc nisky} which created flat-field and sky frames from
the individual observations, respectively, {\sc nireduce} which
subtracted the sky frame from the galaxy images and divided them by
the flat-field frame, and {\sc imcoadd} which aligned and co-added the
dithered galaxy images and removed distortion effects.

\begin{figure}
\includegraphics[width=0.5\textwidth]{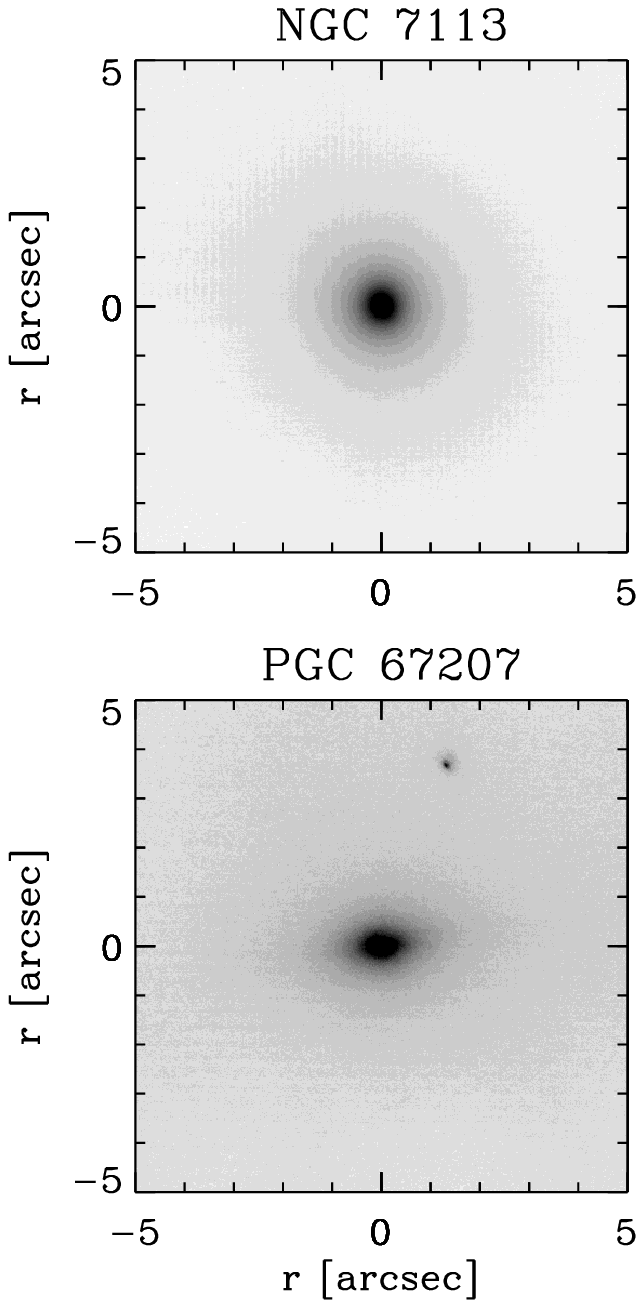}
  \caption{Central portions of the Gemini images of NGC 7113
    (top panel, 1 arcsec = 351.5 pc) and PGC 67207 (bottom
    panel, 1 arcsec = 328.2 pc). Each frame uses a negative
    intensity scale and north is at the top with east to the left.
   \label{fig_images}}
\end{figure}

Fig. \ref{fig_images} shows the final high-resolution images for
NGC~7113 and PGC~67207.

\subsection{Isophotal analysis}
\label{sec_isophotes}

We fitted the isophotes of the sample galaxies in the SDSS and {\em
  Gemini\/} images using the algorithm by \citet{Bender1987}. We
allowed the centres of the ellipses free to vary and considered the
deviation of the isophotal shape from a perfect ellipse to obtain the
radial profiles of the azimuthally-averaged surface brightness, $\mu$,
ellipticity, $\epsilon$, position angle, P.A., centre coordinates,
$x_0$ and $y_0$, and third, fourth, and sixth cosine Fourier
coefficients, $a_3$, $a_4$, and $a_6$, and third, fourth, and sixth
sine Fourier coefficients, $b_3$, $b_4$, and $b_6$.

We derived the $u-g$, $g-r$, $r-i$, and $i-z$ surface-brightness
radial profiles from the SDSS images and we measured the radial
profiles of all the isophotal parameters in the $r$ band for
all the sample galaxies.
We found that the Gemini images are affected by a PSF that
strongly varies with the distance from the FOV centre. Indeed, we
performed a two-dimensional Moffat fit \citep{Moffat1969} to the PSF
standard stars and to the two faint stars visible in the FOV of
NGC~7113 and PGC~67207 at a distance of about 5 arcsec from the galaxy
centre. While the on-axis images of the PSF standard stars were round
($\rm FWHM=0.11$ arcsec, $\beta = 2.4$), the off-axis stellar images
were highly elongated ($1-b/a=0.1$ for the star in the FOV of NGC
7113, 0.2 for the star in the FOV of PGC~67207) and tilted. Therefore,
we derived only the surface-brightness radial profile from the {\em
  Gemini\/} images of NGC~7113 and PGC~67207.

We matched the surface-brightness radial profiles measured in the {\em
  Gemini\/} images to the SDSS ones by determining the zero-point and
sky value of the high-resolution images that minimize the
surface-brightness flux differences with the low-resolution images.
The resulting distribution of the differences $\Delta m = r - K_s$
after minimization gives a mean $\langle \Delta m \rangle = -0.001$
mag and a standard deviation $\sigma_{\Delta m} = 0.012$ mag for
NGC~7113 between 2 and 7 arcsec and $\langle \Delta m \rangle = 0.003$
mag and $\sigma_{\Delta m} = 0.017$ mag for PGC~67207 between 1 and 4
arcsec. We concluded that NGC~7113 and PGC~67207 have no $r-K_s$
  colour gradient in the matched radial region and the scatter
$\sigma_{\Delta m}$ is the dominant source of error in the zero-point
of the Gemini images because the SDSS DR9 $r$ images have a
relative calibration error better than 0.01 mag
\citep{Padmanabhan2008}.
 
The $r$-band radial profiles of the azimuthally averaged surface brightness,
ellipticity, position angle, centre coordinates and third, fourth, and
sixth cosine and sine Fourier coefficients of the sample galaxies are
presented in Fig.~\ref{fig_isophotes}. They are the combination of
the Gemini and SDSS data in the $r$ band for NGC~7113 and
PGC~67207.
The $r$-band SDSS isophotal radial profiles of the sample galaxies are
reported in Table~\ref{tab_photometry_sdss}. The Gemini
surface-brightness radial profiles obtained in the $r$ band for
NGC~7113 and PGC~67207 are given in
Table~\ref{tab_photometry_gemini}. The reported surface brightnesses
are not corrected for cosmological dimming, Galactic absorption, and
$K$ correction.

\subsection{Photometric decomposition}

We derived the structural parameters of NGC~7113 and PGC~67207 by
applying the Galaxy Surface Photometry Two-Dimensional Decomposition
({\sc gasp2d}) algorithm \citep{MendezAbreu2008, MendezAbreu2014} to
the SDSS $r$-band images.

We modelled the surface brightness of both galaxies using a S\'ersic
function \citep{Sersic1968}
%
\begin{equation}
I(x,y) = I_{\rm e} e^{-b_n \left[(r/r_{\rm e})^{1/n}-1\right]},
\end{equation}
%
where $(x,y)$ are coordinates of each image pixel, $r_{\rm e}$ is the
effective radius, $I_{\rm e}$ is the surface brightness at $r_{\rm
  e}$, $n$ is a shape parameter that describes the curvature of the
radial profile, and $b_n = 1.9992\,n-0.3271$ for $0.5<n<10$
\citep{Ciotti1991}. The galaxy model was assumed to have elliptical
isophotes centred on $(x_0,y_0)$ with constant position angle PA and
constant axial ratio $q$. Therefore, the distance $r$ of each image
pixel from the galaxy centre is
%
\begin{eqnarray} 
r(x,y) & = & \left[(-\Delta x \sin{{\rm PA}} 
                   +\Delta y \cos{{\rm PA}})^2 - \right. \nonumber \\  
       &   & \hspace{-0.5cm}\left.(\Delta x \cos{{\rm PA}} 
                   -\Delta y \sin{{\rm PA}})^2/q^2\right]^{1/2}, 
\label{eqn:disc_radius} 
\end{eqnarray} 
%
where $\Delta x = x-x_0$ and $\Delta y = y-y_0$. 
The total luminosity of the model galaxy is 
%
\begin{equation}
L_{\rm T} = 2\,\pi\,n\,(1-q)\,I_{\rm e}\,r_{\rm e}^2\,e^{b_n}\,{b_n^{-2n}}\,\Gamma(2n)
\end{equation}
%
where $\Gamma$ is the complete gamma function.

To obtain the best-fitting structural parameters $I_{\rm e}$, $r_{\rm
  e}$, $n$, PA, and $q$ and the position of the galaxy centre $(x_0,
y_0)$, we took into account the seeing effects by convolving the model
image with a circular Moffat PSF with the shape parameters measured
directly from stars in the galaxy images. We found $\rm FWHM=0.96$
arcsec and $\beta = 4.0$ for NGC~7113 and $\rm FWHM=0.99$ arcsec and
$\beta = 2.7$ for PGC~67207. We fitted both images out to a surface
brightness level of $1.5\sigma_{\rm sky}$ corresponding to
$\mu_r=24.1$ mag arcsec$^{-2}$. We adopted the same masks built for
fitting the isophotes and excluded the masked pixels from the fit.

We estimated the errors on the best-fitting parameters by generating a
set of 200 images of galaxies with a S\'ersic surface-brightness
radial profile and total apparent magnitude $12\leq\,m_r\,\leq14$
mag. We randomly chose the structural parameters of the artificial
galaxies to cover the ranges obtained for our galaxies. We adopted the
values of the pixel scale (0.396 arcsec pixel$^{-1}$), gain (4.8 $e^-$
ADU$^{-1}$), readout noise (5.6 $e^-$ rms), and exposure time ($53.9$
s) to mimic the instrumental setup of the photometric observations. We
added to the artificial images a background level (120 ADU) and photon
noise to yield a signal-to-noise ratio (S/N) similar to that of the
observed ones. We adopted a Moffat PSF with $\rm FWHM=0.98$ arcsec and
$\beta = 3.3$. Finally, we analysed the images of the artificial
galaxies with {\sc gasp2d} and calculated the relative errors on the
fitted parameters. We derived the mean and standard deviation of the
relative errors for the artificial galaxies of each magnitude bin and
we adopted them as the systematic and statistical errors for the
best-fitting parameters of the observed galaxies according to their
apparent magnitude.

Fig.~\ref{fig_GASP2D} shows the $r$-band SDSS image, {\sc gasp2d}
best-fitting image, and residual image of NGC~7113 and PGC~67207,
respectively. Their best-fitting structural parameters are collected
in Table~\ref{tab_fit}. The effective surface brightnesses are
corrected for cosmological dimming, Galactic absorption, and $K$
correction and the total magnitudes of the S\'ersic models are given
after applying the Galactic absorption and $K$ corrections. For both
galaxies, we found that the best-fitting S\'ersic function is a good
representation of the combined Gemini-SDSS surface-brightness
radial profile (Fig. \ref{fig_isophotes}). Therefore, we adopted such
a parametric representation for the dynamical modelling of the two
galaxies.

\begin{figure*}
  \begin{center}
    \includegraphics[scale=0.9]{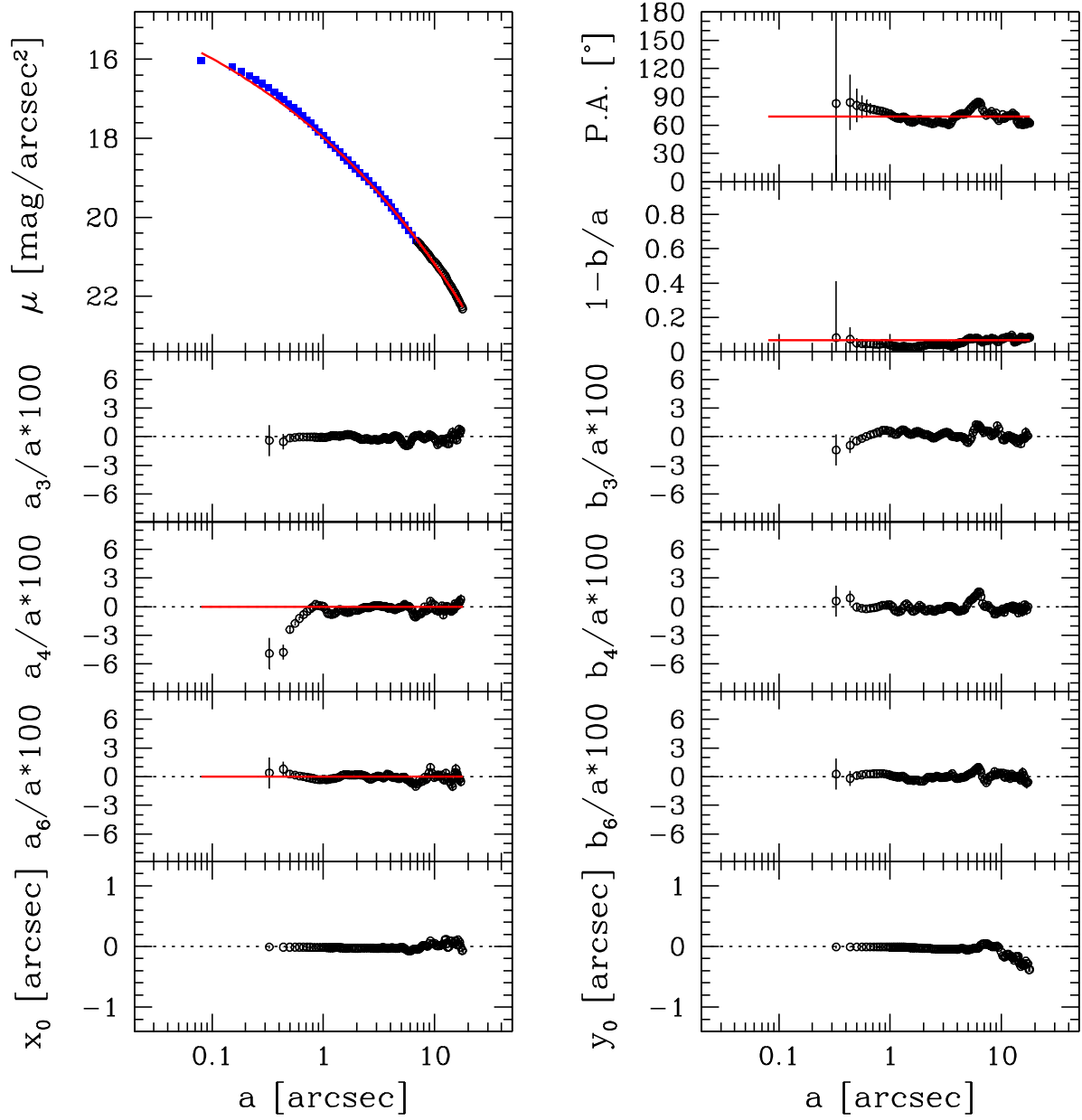}
  \end{center}
  \caption{Isophotal parameters of the galaxies from $r$-band SDSS
    (black open circles) and Gemini images (blue filled
    squares) as a function of the logarithm of the semimajor-axis
    distance in arcsec. Left-hand panels (from top to bottom): radial
    profiles of surface brightness ($\mu$), third- ($a_3$), fourth-
    ($a_4$), and sixth-order ($a_6$) cosine Fourier coefficients, and
    $x$-coordinate of the centre ($x_0$). Right-hand panels (from top
    to bottom): radial profiles of position angle (PA), ellipticity
    ($e=1-b/a$), third- ($b_3$), fourth- ($b_4$), and sixth-order
    ($b_6$) sine Fourier coefficients, and $y$-coordinate of the
    centre ($y_0$).  For NGC~7113 and PGC~67207 the solid red lines
    correspond to the radial profiles of $\mu$, PA, $e$, $a_4$, and
    $a_6$ without the correction for seeing convolution which we
    obtained from the photometric decomposition and used in the
    dynamical modelling.  The measured and model surface brightnesses
    are not corrected for cosmological dimming, Galactic absorption,
    and $K$ correction.
\label{fig_isophotes}}

\end{figure*}
\addtocounter{figure}{-1}
\begin{figure*}
  \begin{center}
    \includegraphics[scale=0.9]{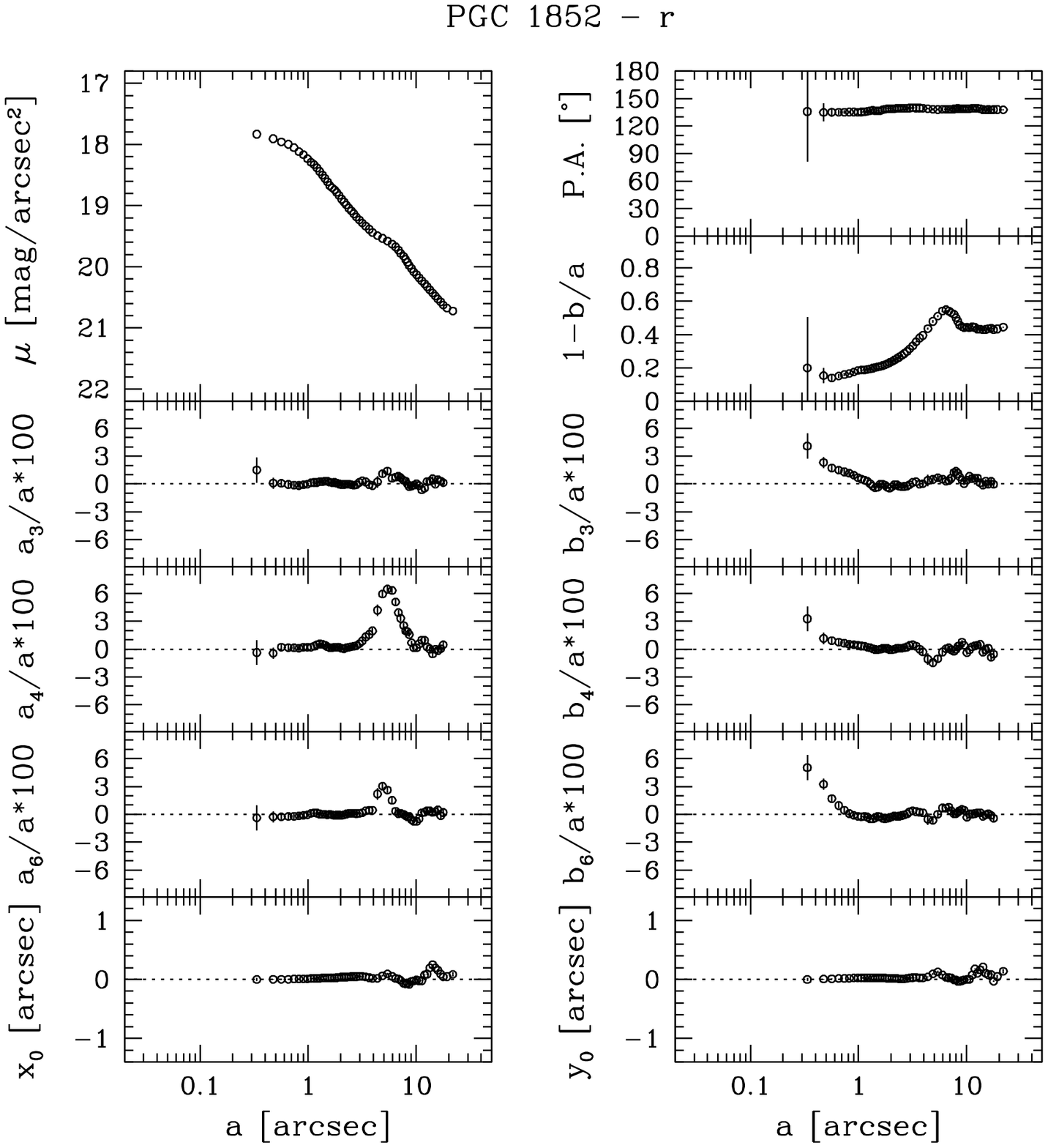}
  \end{center}
  \caption{{\em continued}}
\end{figure*}

\begin{figure*}
\addtocounter{figure}{-1}
  \begin{center}
    \includegraphics[scale=0.9]{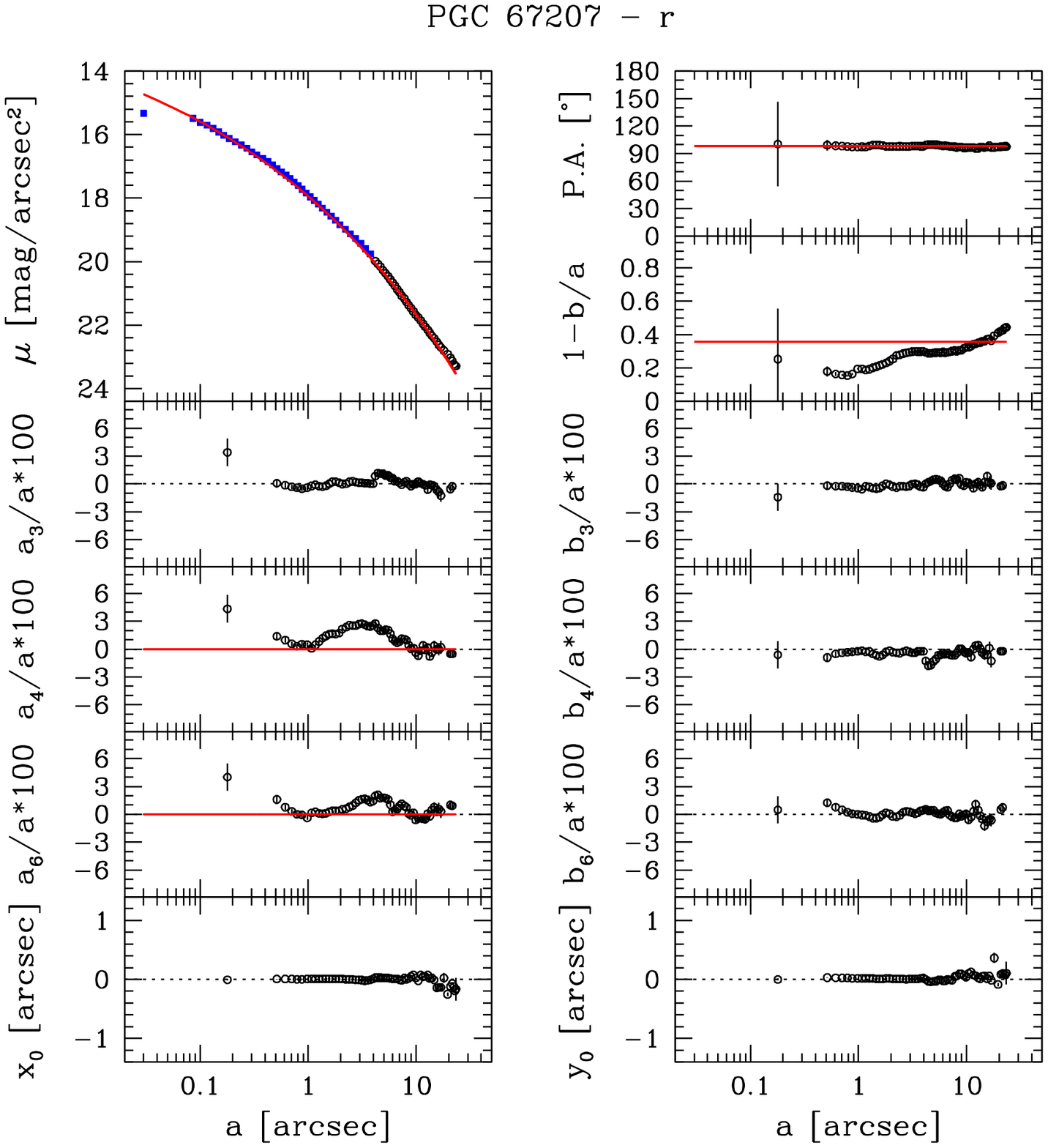}
  \end{center}
  \caption{{\em continued}}
\end{figure*}

\begin{figure*}
  \begin{center}
\begin{tabular}{c}
\vspace{4.5cm}
\includegraphics[angle=90,scale=.7,bb=558 55 350 737]{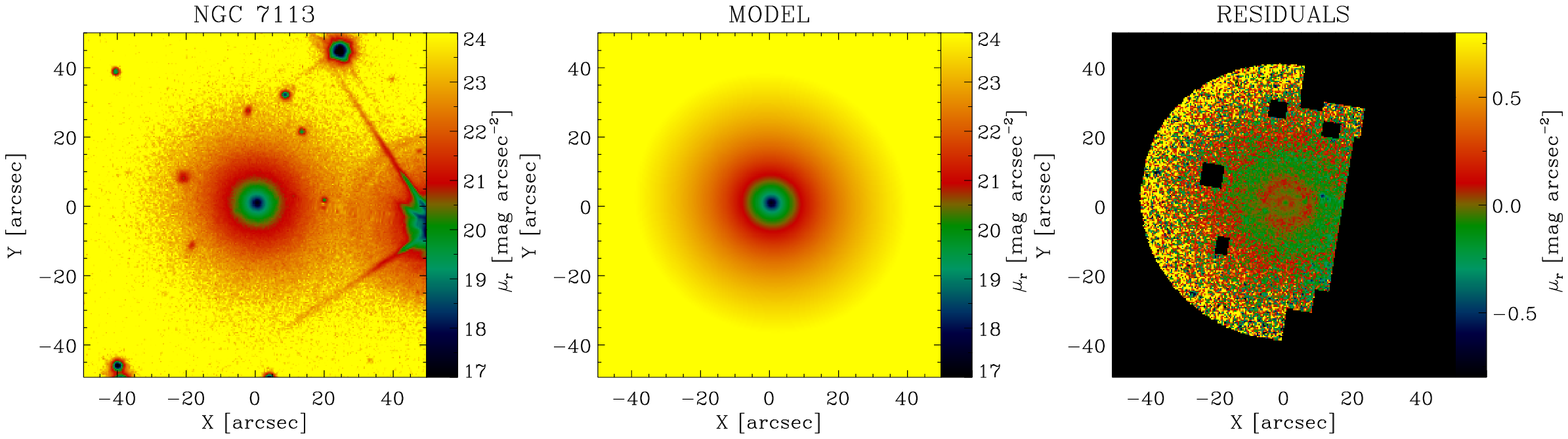}\\
\includegraphics[angle=90,scale=.7,bb=558 55 350 737]{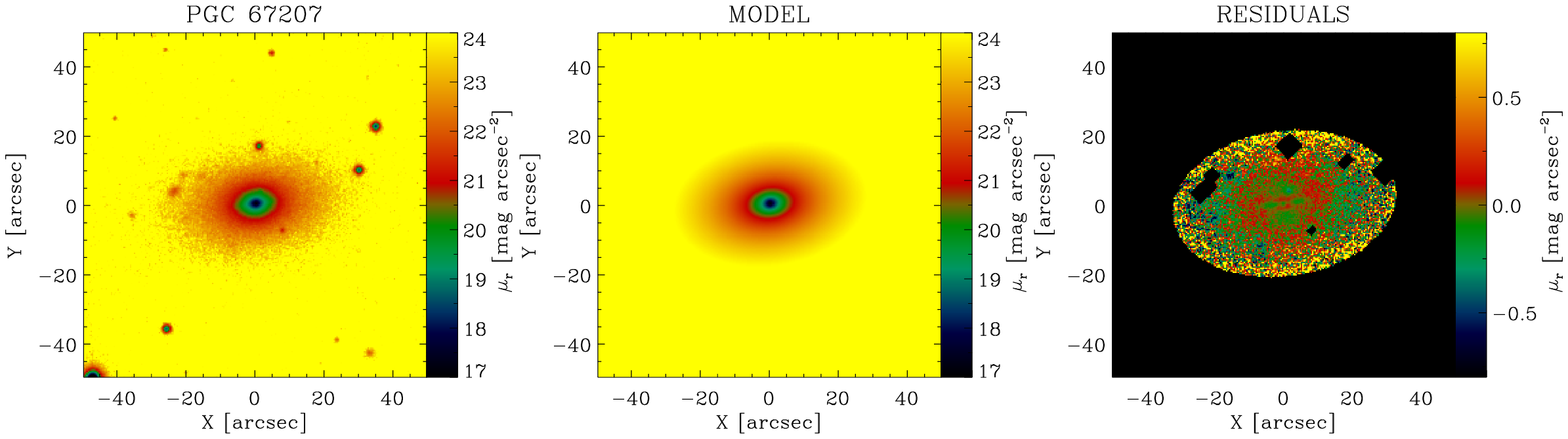}
\end{tabular}

\vspace{5cm}

  \end{center}
  \caption{Two-dimensional photometric decomposition of NGC~7113 (top
    panels) and PGC~67207 (bottom panels). The SDSS $r$-band image
    (left-hand panel), best-fitting image (middle panel), and residual
    (i.e., observed-model) image (right-hand panel) are shown. The
    black areas in the residual images correspond to pixels excluded
    from the fitting procedure. \label{fig_GASP2D}}
\end{figure*}

\begin{table*}  
\begin{center}
\begin{small}
\caption{The structural parameters of NGC~7113 and PGC~67207 from the
  photometric decomposition.\label{tab_fit}}
\begin{tabular}{lccccccc}
\noalign{\smallskip}   
\hline
\multicolumn{1}{c}{Object} & 
\multicolumn{1}{c}{$\mu_{{\rm eff},r}$} &
\multicolumn{1}{c}{$r_{\rm eff}$} &
\multicolumn{1}{c}{$n$} & 
\multicolumn{1}{c}{$q$} & 
\multicolumn{1}{c}{PA} &
\multicolumn{1}{c}{$m_r$} & 
\multicolumn{1}{c}{$M_r$} \\
\multicolumn{1}{c}{} & 
\multicolumn{1}{c}{(mag arcsec$^{-2}$)} &
\multicolumn{1}{c}{(arcsec)} & 
\multicolumn{1}{c}{} & 
\multicolumn{1}{c}{} & 
\multicolumn{1}{c}{($^\circ$)} &
\multicolumn{1}{c}{(mag)} & 
\multicolumn{1}{c}{(mag)} \\ 
\multicolumn{1}{c}{(1)} & 
\multicolumn{1}{c}{(2)} &   
\multicolumn{1}{c}{(3)} & 
\multicolumn{1}{c}{(4)} & 
\multicolumn{1}{c}{(5)} & 
\multicolumn{1}{c}{(6)} &
\multicolumn{1}{c}{(7)} &  
\multicolumn{1}{c}{(8)} \\   
\noalign{\smallskip}   
\hline
\noalign{\smallskip}       
 NGC~7113 & $22.54\pm0.02$ & $24.83\pm12.50$& $4.89\pm0.47$ & $0.932\pm0.003$ & $69.1\pm0.1$ & 12.23 & $-22.15$\\ 
PGC~67207 & $21.71\pm0.02$ & $12.44\pm6.26$ & $5.18\pm0.50$ & $0.642\pm0.003$ & $97.8\pm0.2$ & 13.27 & $-20.96$\\ 
\noalign{\smallskip}       
\hline
\end{tabular} 
\end{small}
\end{center}     
\begin{minipage}{17cm}
{\em Note.} Column (1): galaxy name. Column (2): the SDSS $r$-band
effective surface brightness corrected for cosmological dimming and
after applying the Galactic absorption \citep{Schlafly2011} and $K_r$
corrections \citep{Chilingarian2010} available in NED. Column (3):
effective radius. Column (4): S\'ersic shape parameter. Column (5):
axial ratio. Column (6): major-axis position angle measured North
through East. Column (7): total apparent magnitude of the S\'ersic
model after applying the Galactic absorption \citep{Schlafly2011} and
$K_r$ corrections \citep{Chilingarian2010} available in NED. Column
(8): total absolute magnitude adopting the distance modulus given in
Table~\ref{tab_galaxies}. 
\end{minipage}
\end{table*}   

\begin{table*}  
\caption{Instrumental set up of the spectroscopic observations.\label{tab_setup}}     
\begin{center}
\begin{small}
\begin{tabular}{lcccc}    
\hline 
\noalign{\smallskip}   
\multicolumn{1}{c}{Parameter} & 
\multicolumn{1}{c}{Run 1} &
\multicolumn{1}{c}{Run 2} &  
\multicolumn{1}{c}{Run 3} & 
\multicolumn{1}{c}{Run 4} \\   
\multicolumn{1}{c}{(1)} & 
\multicolumn{1}{c}{(2)} &   
\multicolumn{1}{c}{(3)} & 
\multicolumn{1}{c}{(4)} &
\multicolumn{1}{c}{(5)} \\   
\noalign{\smallskip}   
\hline
\noalign{\smallskip}       
Spectrograph & \multicolumn{2}{c}{Moderate resolution} & Boller and Chivens & Moderate resolution \\ 
Grating (gr~mm$^{-1}$) & \multicolumn{2}{c}{1200} & 600 & 1200 \\
Detector name  & \multicolumn{2}{c}{Echelle} & Echelle Ohio State University Loral C & Templeton \\
Detector manufacturer & \multicolumn{2}{c}{SITe} & Loral & SITe \\
Pixel number & \multicolumn{2}{c}{$2048\times2048$} &  $1200\times800$ & $1024\times1024$ \\  
Pixel size  ($\mu$m$^2$) & \multicolumn{2}{c}{$24\times24$} & $15\times15$ & $24\times24$ \\
Gain ($e^-$ ADU$^{-1}$) & \multicolumn{2}{c}{2.7} & 2.1 & 3.5 \\
Read-out noise ($e^-$ rms) & \multicolumn{2}{c}{7.9} & 7.0 & 5.3 \\
Scale (arcsec pixel$^{-1}$) & \multicolumn{2}{c}{0.606} & 0.41 & 0.28 \\ 
Dispersion (\AA\ pixel$^{-1}$) & 0.870 & 0.906 & 0.751 & 0.923 \\
Slit width (arcsec) & \multicolumn{2}{c}{1.9} & 1.7 & 1.9 \\
Wavelength range (\AA) & 4670-6451 & 4675-6530 & 4250-5151 & 4700-5645 \\
Instrumental FWHM (\AA) & 2.31 & 2.40 & 3.05 & 1.68 \\
Instrumental $\sigma$ at \Hb\ (km s$^{-1}$) & 60 & 63 & 80 & 44 \\
\noalign{\smallskip}       
\hline
\end{tabular} 
\end{small}
\end{center}     
\end{table*}    

\section{Long-slit spectroscopy}
\label{sec_spectroscopy}

\subsection{Observations and data reduction}
\label{sec_obsred}

Long-slit spectroscopic data of the sample galaxies were obtained with
the 2.4-m Hiltner telescope of the MDM Observatory at Kitt Peak,
Arizona, during four different runs between 2003 and 2012. Details of
the instrumental set-up of the observations carried out on 2003
November 18-21 (run 1), 2005 November 01 (run 2), 2006 November 13-16
(run 3), and 2012 September 24-25 (run 4) are given in
Table~\ref{tab_setup}.

The sample galaxies were observed along different axes crossing the
nucleus. At the beginning of each exposure the galaxy was centred on
the slit using the guiding camera. NGC~7113 was observed in two
different runs at position angle $\pa=0^\circ$ to perform a
consistency check between the measurements of stellar kinematics and
line-strength indices. The integration times of each exposure, total
integration times, and slit position angle of the galaxy spectra are
given in Table~\ref{tab_log}.
 
In each run, a number of G and K giant stars were selected from the
samples by \citet{Faber1985} and \citet{Gonzalez1993} and observed to
use their spectra as templates in measuring the line-strength
indices. In addition, different spectra of at least one
spectrophotometric standard star per night were obtained to calibrate
the flux of the galaxy and template star spectra before line-strength
indices were measured. A spectrum of the comparison arc lamp was taken
before and/or after each object exposure to allow an accurate
wavelength calibration. Bias exposures and quartz-lamp flat-field
spectra were acquired in the afternoon before each observing night.
 
We performed the reduction of the spectroscopic data using standard
{\sc iraf} routines. We bias subtracted, flat-field corrected,
corrected for cosmic rays and bad columns, wavelength
calibrated, corrected for CCD misalignment, sky subtracted, and flux
calibrated all the spectra. No pixel binning was adopted.
After bias subtraction, we corrected the spectra for pixel-to-pixel
sensitivity variations by using an averaged and normalized flat-field
for each night.  We identified and corrected cosmic rays by
interpolating over with the {\sc lacos\_spec} task
\citep{vanDokkum2001}. We corrected the residual cosmic rays by
manually editing the spectra. Bad columns were replaced by linearly
interpolating over adjacent columns. We rebinned each spectrum using
the wavelength solution obtained from the corresponding arc-lamp
spectrum. We checked the accuracy of the wavelength rebinning by
measuring the rms of the difference between the derived and predicted
wavelengths for the arc-lamp emission lines in the
wavelength-calibrated comparison spectra. For all the runs the
resulting rms is $0.05$ \AA\ corresponding to $3$ \kms\ at \Hb , while
the systematic error of the wavelength calibration is smaller than
$10$ \kms. The latter was estimated from the wavelength of the
night-sky emission lines which we identified in the galaxy spectra
\citep{Osterbrock1996}. The instrumental resolution of each run is
given in Table~\ref{tab_setup}. It was derived as the mean of the
Gaussian FWHMs measured for the unblended arc-lamp emission lines of
the wavelength-calibrated comparison spectra. We corrected all the
galaxy and star spectra for CCD misalignment and determined the sky
contribution by interpolating a one-degree polynomial along the
outermost 20 arcsec at the two edges of the slit, where the galaxy or
stellar light was negligible. The estimated sky level was subtracted
from the spectra. We flux calibrated each galaxy and template star
spectrum using the sensitivity function obtained from the spectra of
the spectrophotometric standard star obtained in the corresponding
night. Finally, we co-added the spectra obtained for the same galaxy
along the same axis in each run by using the centre of the stellar
continuum as reference. This allowed us to improve the S/N of the
resulting two-dimensional spectrum. A one-dimensional spectrum was
obtained for each template star and we deredshifted it to rest frame.

\renewcommand{\tabcolsep}{3pt}
\begin{table}  
\caption{Log of the spectroscopic observations. \label{tab_log}}
\begin{center}
\begin{small}
\begin{tabular}{lcrccc}    
\hline 
\noalign{\smallskip}   
\multicolumn{1}{c}{Galaxy} & \multicolumn{1}{c}{Run} &  
\multicolumn{1}{c}{PA} & \multicolumn{1}{c}{Position} &   
\multicolumn{1}{c}{Single Exp. T.} & \multicolumn{1}{c}{Total Exp. T.} \\   
\multicolumn{1}{c}{} & \multicolumn{1}{c}{} &  
\multicolumn{1}{c}{($^\circ$)} & \multicolumn{1}{c}{} &  
\multicolumn{1}{c}{(s)} & \multicolumn{1}{c}{(h)} \\   
\multicolumn{1}{c}{(1)} & \multicolumn{1}{c}{(2)} &   
\multicolumn{1}{c}{(3)} & \multicolumn{1}{c}{(4)} &   
\multicolumn{1}{c}{(5)} & \multicolumn{1}{c}{(6)} \\ 
\noalign{\smallskip}   
\hline
\noalign{\smallskip}       
NGC\,7113     & 1 &     0 & ... & $3\times3600$ & 3.0 \\ 
              & 3 &     0 & ... & $2\times1800$ & 1.0 \\ 
              & 4 &    90 & ... & $6\times3600$ & 6.0 \\ 
PGC\,1852     & 1 & $-42$ & MJ  & $3\times3600$ & 3.0 \\ 
              & 2 &    48 & MN  & $3\times3600$ & 3.0 \\ 
PGC\,67207    & 1 & $-83$ & MJ  & $2\times3600$ & 2.0 \\ 
              & 3 &     0 & DG  & $2\times1800$ & 1.0 \\ 
              & 4 &     7 & MN  & $5\times3600$ & 5.0 \\ 
\noalign{\smallskip}       
\hline
\end{tabular} 
\end{small}
\end{center}     
\begin{minipage}{8.5cm}
{\em Note.} Column (1): galaxy name. Column (2): observing run. Column (3):
slit position angle measured North through East. Column (4): slit
position: MJ = major axis, MN = minor axis, DG = diagonal axis. Column (5):
number and exposure time of the single exposures. Column (6): total exposure
time.
\end{minipage}
\end{table}    

\subsection{Stellar kinematics}
\label{sec_kinematics}

We measured the line-of-sight velocity distribution (LOSVD) of the
stellar component of the sample galaxies from the absorption lines in
the observed wavelength ranges using the Penalized Pixel Fitting
\citep[{\sc ppxf};][]{Cappellari2004} and Gas and Absorption Line
Fitting \citep[{\sc gandalf};][]{Sarzi2006} algorithms which we
adapted to deal with MDM spectra. The LOSVD was assumed to be a
Gaussian plus third- and fourth-order Gauss-Hermite polynomials
\citep{Gerhard1993, vanderMarel1993}.
 
We rebinned each galaxy spectrum along the dispersion direction to a
logarithmic scale, and along the spatial direction to obtain an
$\rm S/N\geq20$ per resolution element. For each radial bin, we built an
optimal template spectrum by convolving a linear combination of the
stellar spectra available in the Medium Resolution Isaac Newton
Telescope Library of Empirical Spectra
\citep[MILES;][]{Sanchez-Blazquez2006, Falcon-Barroso2011} with the
LOSVD in order to fit the galaxy spectrum. The optimal template
spectrum and LOSVD moments were obtained by $\chi^2$ minimization in
pixel space. Before fitting, the MILES stellar spectra were
logarithmically rebinned and dereshifted to rest frame. Moreover, we
degraded the spectral resolution of the galaxy spectrum by convolving
it with a Gaussian function in order to match the MILES spectral
resolution \citep[$\rm FWHM=2.54$ \AA;][]{Beifiori2011}. In addition,
we simultaneously fitted the ionized-gas emission lines detected with
a $\rm S/N>3$.  We masked the bad pixels coming from imperfect subtraction
of cosmic rays and sky emission lines and excluded them from the
fitting procedure. We added a low-order multiplicative Legendre
polynomial to correct for the different shape of the continuum in the
spectra of the galaxy and optimal template due to reddening and
large-scale residuals of flat-fielding and sky subtraction.

By measuring the LOSVD moments in all the available radial bins along
the spatial direction we derived the radial profiles of the
line-of-sight velocity $v$, velocity dispersion $\sigma$, third-order
Gauss-Hermite moment $H_3$, and fourth-order Gauss-Hermite moment
$H_4$ of the stars. We estimated the uncertainties on the LOSVD
moments running Monte Carlo simulations. For each radial bin we built
a set of simulated galaxy spectra by randomly perturbing the
best-fitting galaxy spectrum. We added to the counts of each pixel of
the best-fitting galaxy spectrum a random value chosen from a Gaussian
distribution with a mean of zero and the same standard deviation of
the difference between the observed and best-fitting galaxy spectra in
the wavelength range used in the fit and excluding the emission
lines. We measured the simulated spectra as if they were real. For
each LOSVD moment we adopted as error the standard deviation of the
distribution of the values derived for the simulated galaxy
spectra. We found no bias of the {\sc ppxf} method with the adopted
instrumental setups and spectral samplings in the ranges of S/N and
$\sigma$ which characterize the spectra of the sample galaxies. Indeed
the values of $H_3$ and $H_4$ measured in a set of simulated galaxy
spectra obtained by convolving the best-fitting galaxy spectrum with a
Gauss-Hermite LOSVD, and adding photon, readout, and sky noise to
mimic actual observations differ from the intrinsic ones only within
the estimated errors. The general trends and features of the LOSVD
moments measured from the multiple observations of NGC~7113 agree
within the errors.
 
The measured stellar kinematics are reported in
Table~\ref{tab_kinematics} where velocities are relative to the galaxy
centres. The folded kinematic profiles are plotted in the left-hand
panels of Fig.~\ref{fig_kinematics}.

\begin{figure*}
  \includegraphics[width=9.9cm]{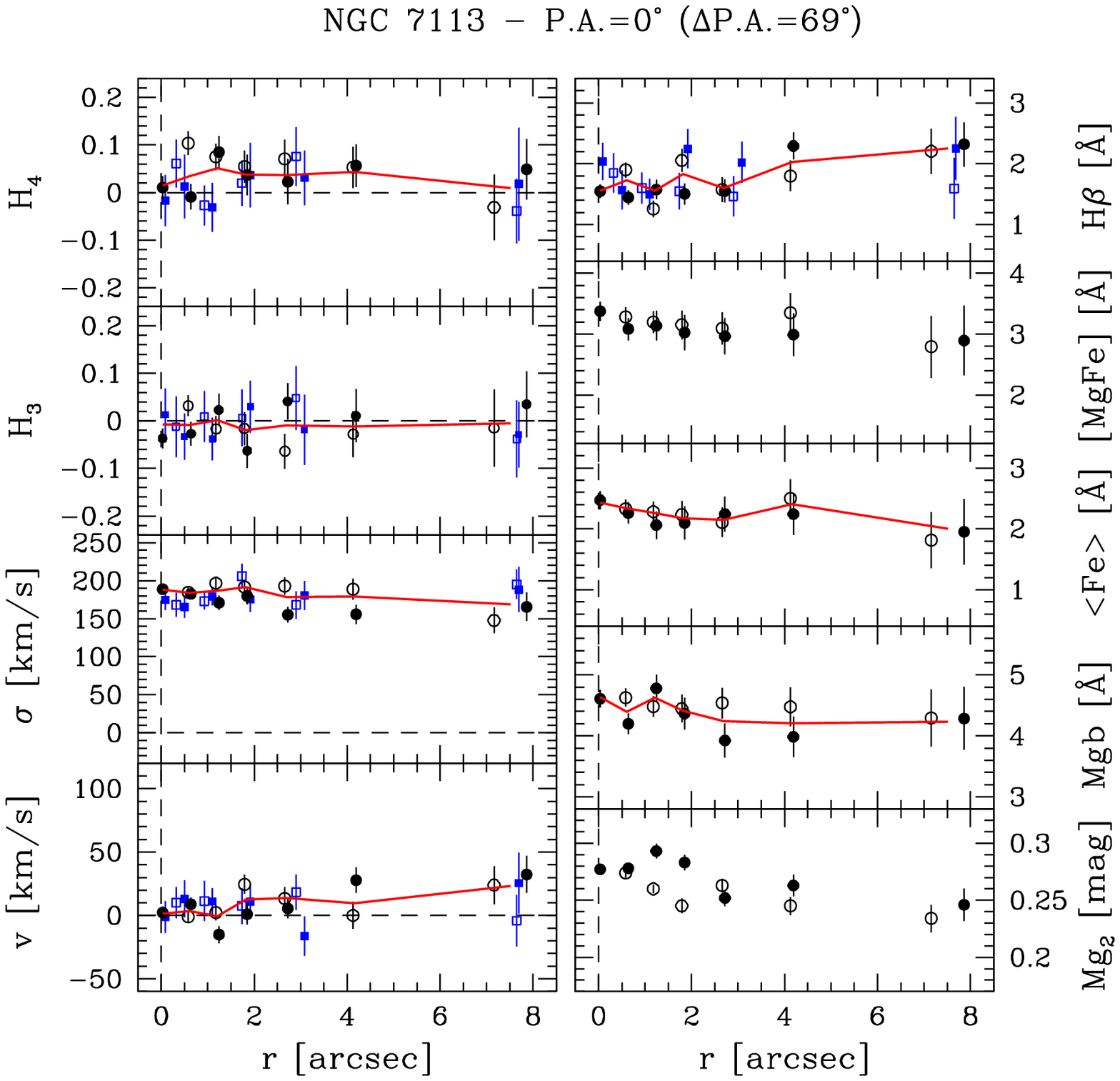}
  \includegraphics[width=5.7cm]{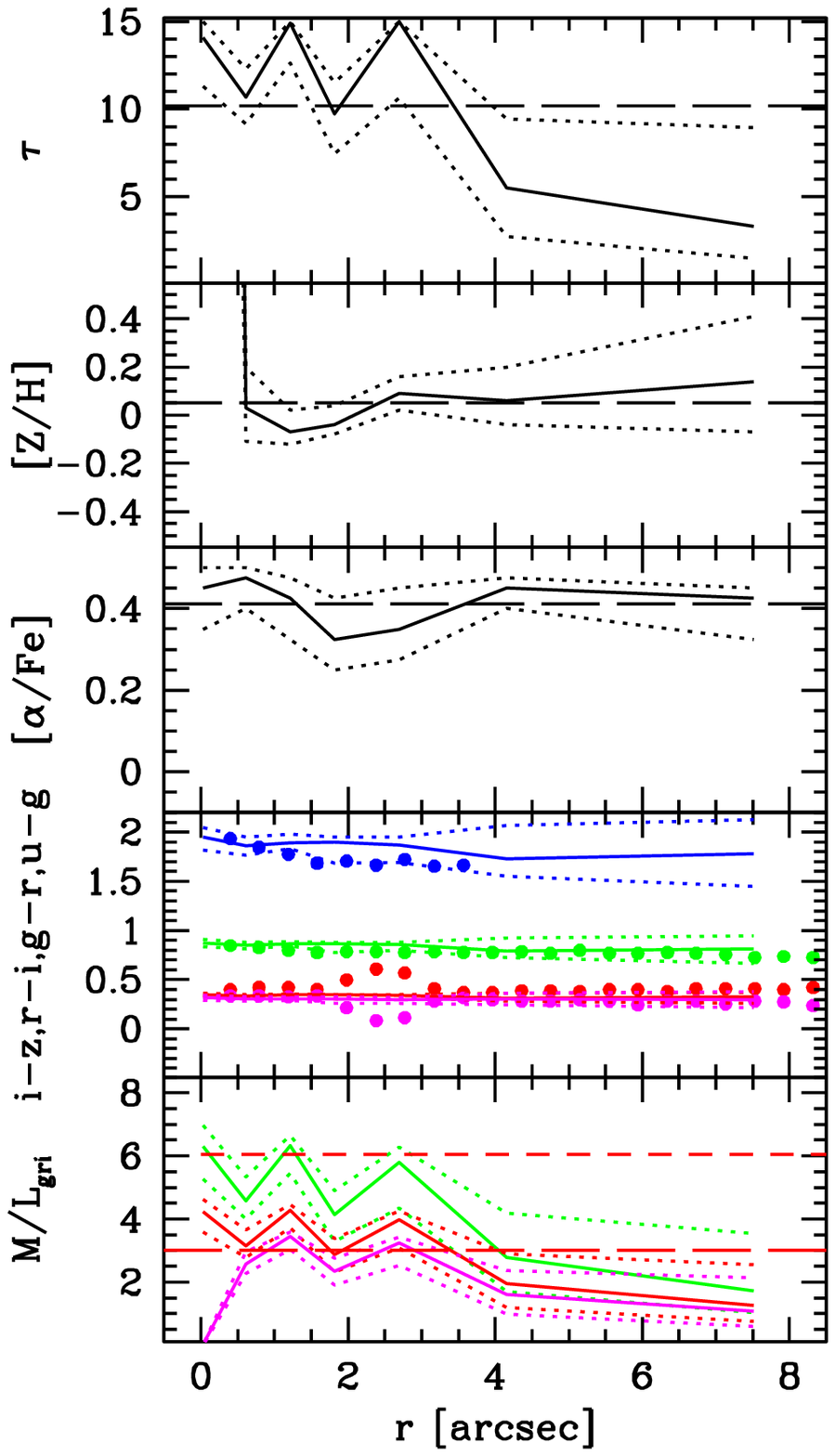}
  \caption{Folded stellar kinematics, line-strength indices, and
    stellar population parameters along the observed axes of the
    sample galaxies.
    Left-hand panels (from bottom to top): radial profiles of the LOS
    velocity ($v$) after the subtraction of systemic velocity,
    velocity dispersion ($\sigma$), third- ($H_3$), and fourth-order
    ($H_4$) coefficient of the Gauss-Hermite decomposition of the
    LOSVD. The black circles are for data measured in runs 1, 2 and 4,
    while blue squares are for run 3. Filled and open symbols refer to
    data measured along the receding and approaching side,
    respectively. For NGC 7113 and PGC 67207 the red solid lines
    correspond to the stellar kinematic parameters of the best-fitting
    dynamical model. The difference ($\Delta$PA) between the PAs
    of the observed axis and the line of nodes adopted in the
    dynamical model is also given.
    Middle panels (from top to bottom): radial profiles of the \Hb ,
    \MgFe , \Fe , \Mgb, and \Mgd\ line-strength indices. The red solid
    lines correspond to the line-strength indices derived from the
    best-fitting SSP models with a Kroupa IMF. Symbols are the same as
    in the left-hand panels.
    Right-hand panels (from top to bottom): Radial profiles of
    stellar-population age (\age), metallicity (\ZH), and
    $\alpha$-elements overabundance (\aFe) derived from the
    best-fitting SSP models with a Kroupa IMF with long-dashed lines
    marking their mean values. Below them the predicted $u-g$ (blue
    solid line), $g-r$ (green solid line), $r-i$ (red solid line) and
    $i-z$ (magenta solid line) colours are shown together with the
    measured ones (blue, green, red, and magenta filled circles). At
    the bottom, the predicted mass-to-light ratios in the $g$ (green
    solid line), $r$ (red solid line), and $i$ (magenta solid line)
    band are given with the long-dashed line marking to the mean value
    of the mass-to-light ratio in the $r$-band $\mlkroupa$. For NGC
    7113 and PGC 67207, the short-dashed line corresponds to the
    dynamical mass-to-light ratio $\mldyn$. The dotted lines show the
    error ranges for all the plotted quantities.
\label{fig_kinematics}}
\end{figure*}

\addtocounter{figure}{-1}
\begin{figure*}
  \includegraphics[width=9.9cm]{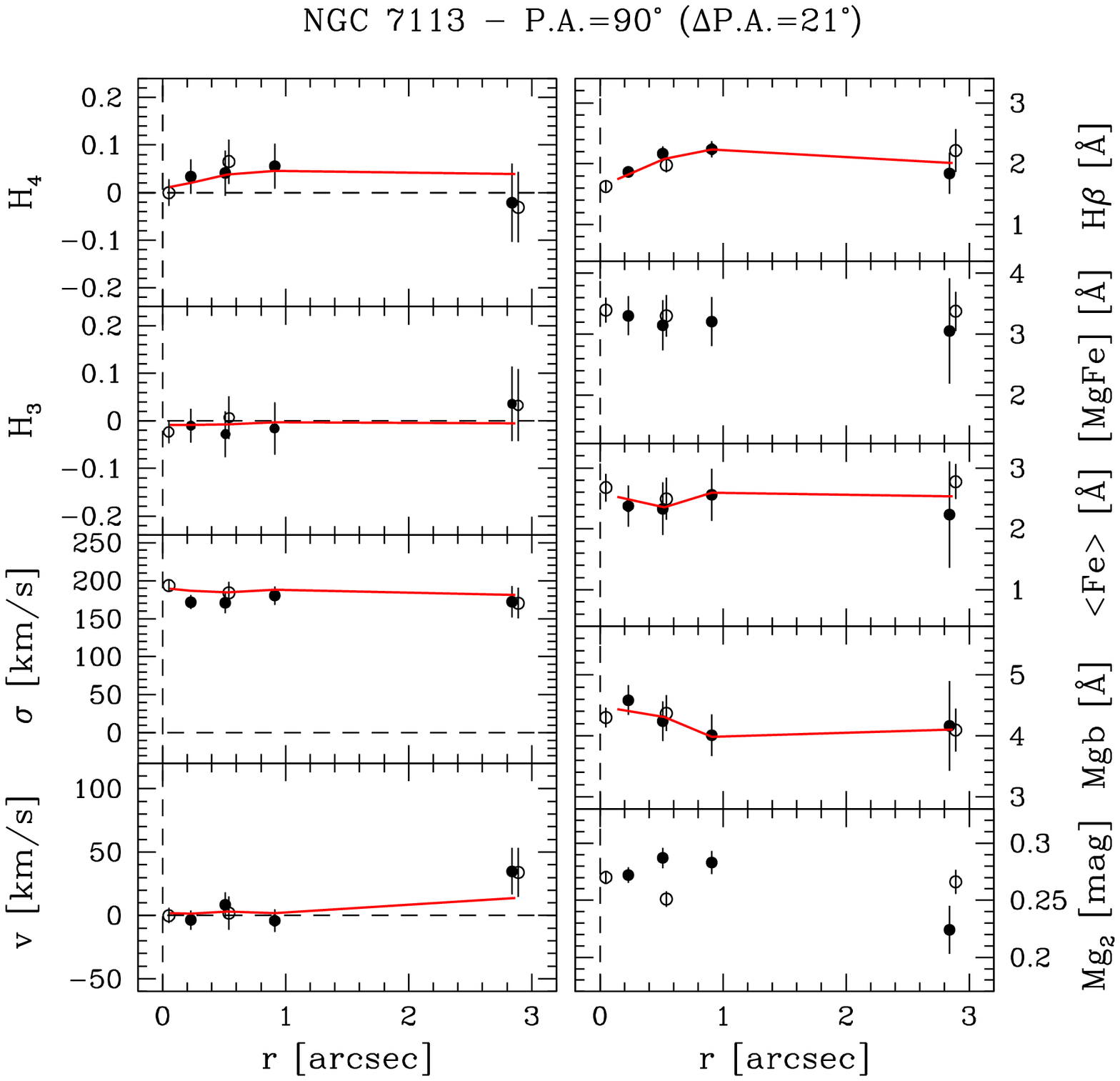}
  \includegraphics[width=5.7cm]{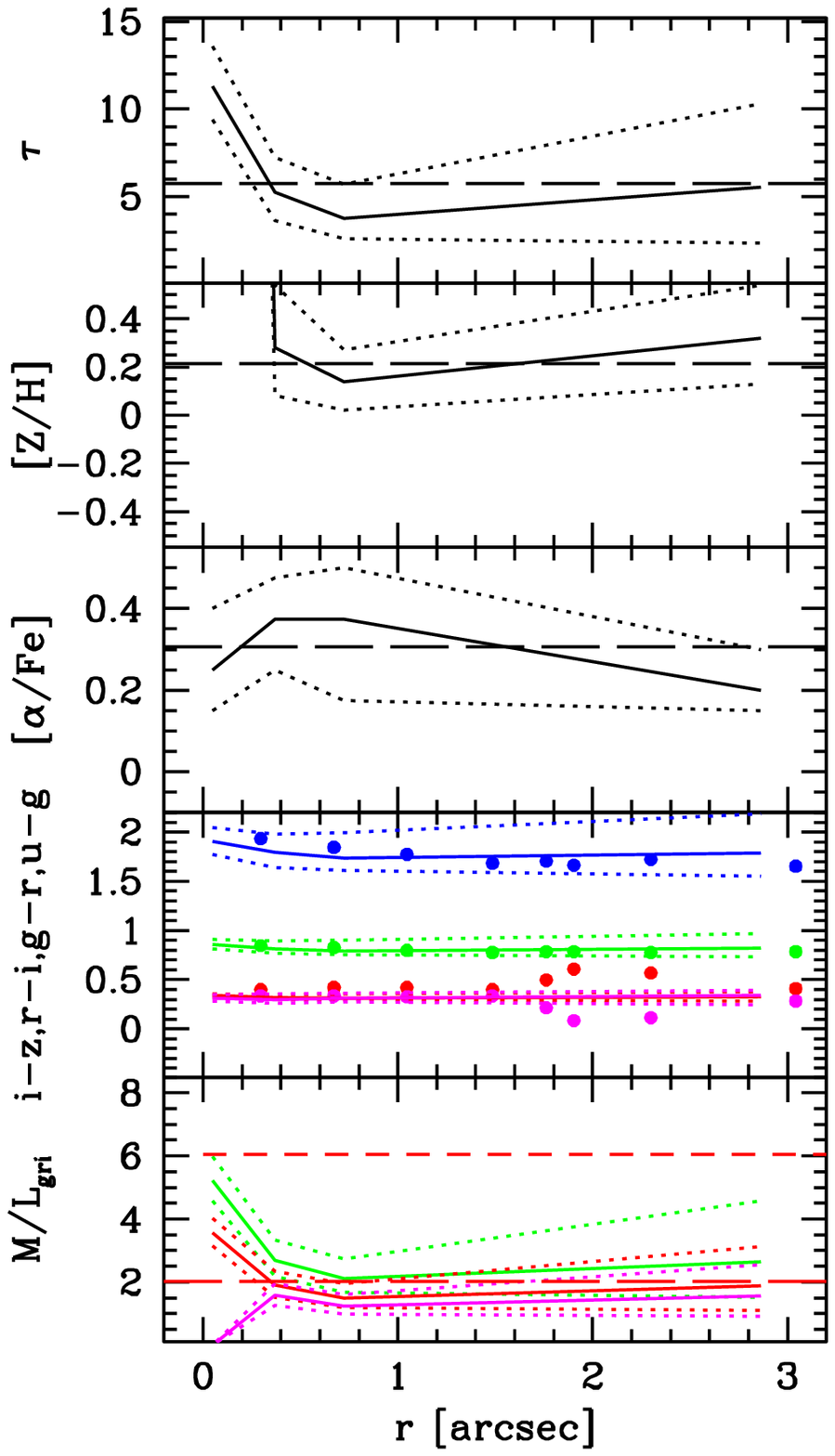}

   \includegraphics[width=9.9cm]{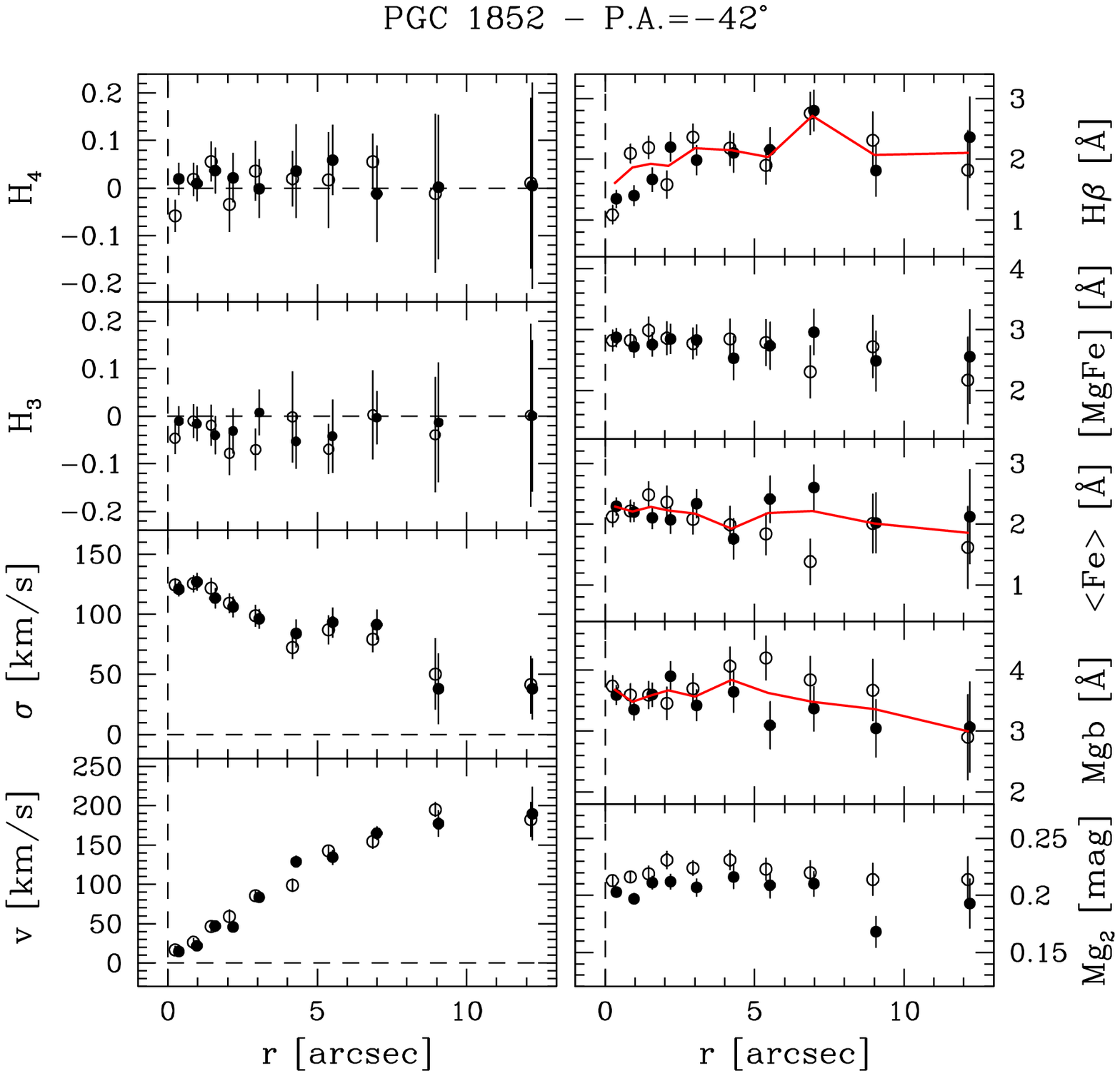} 
   \includegraphics[width=5.7cm]{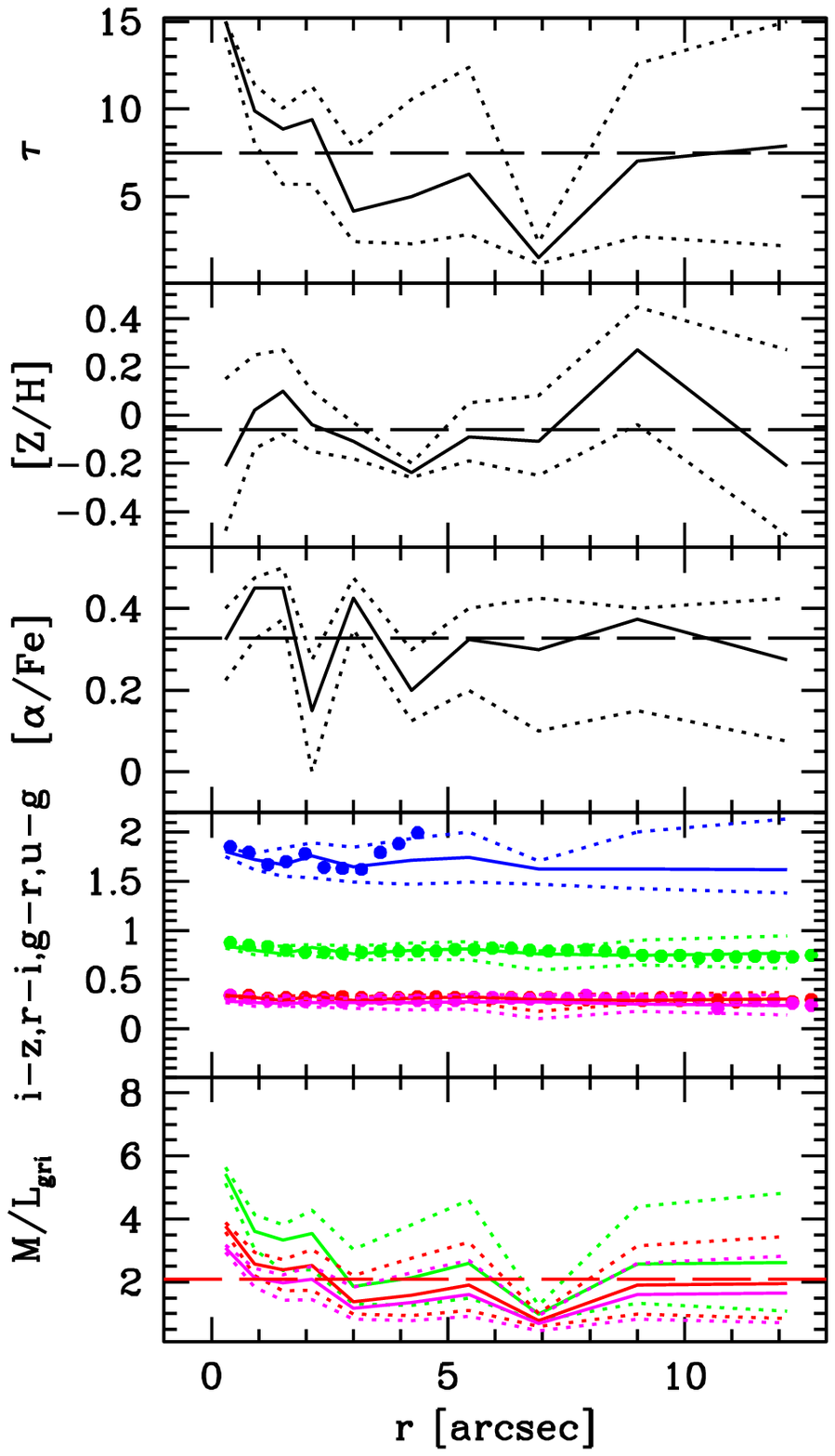}
  \caption{{\em continued}}
\end{figure*}

\addtocounter{figure}{-1}
\begin{figure*}
  \includegraphics[width=9.9cm]{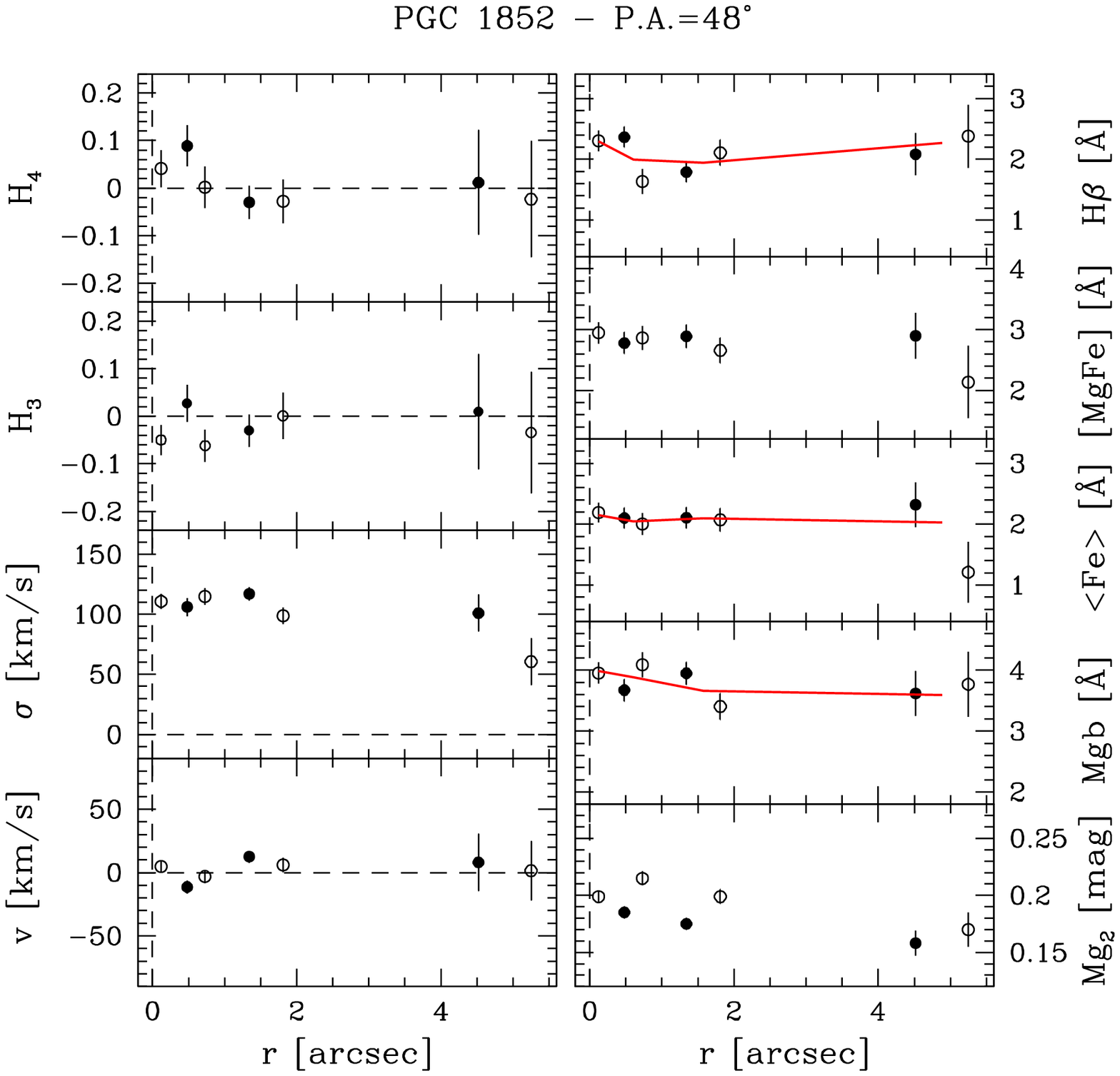} 
  \includegraphics[width=5.7cm]{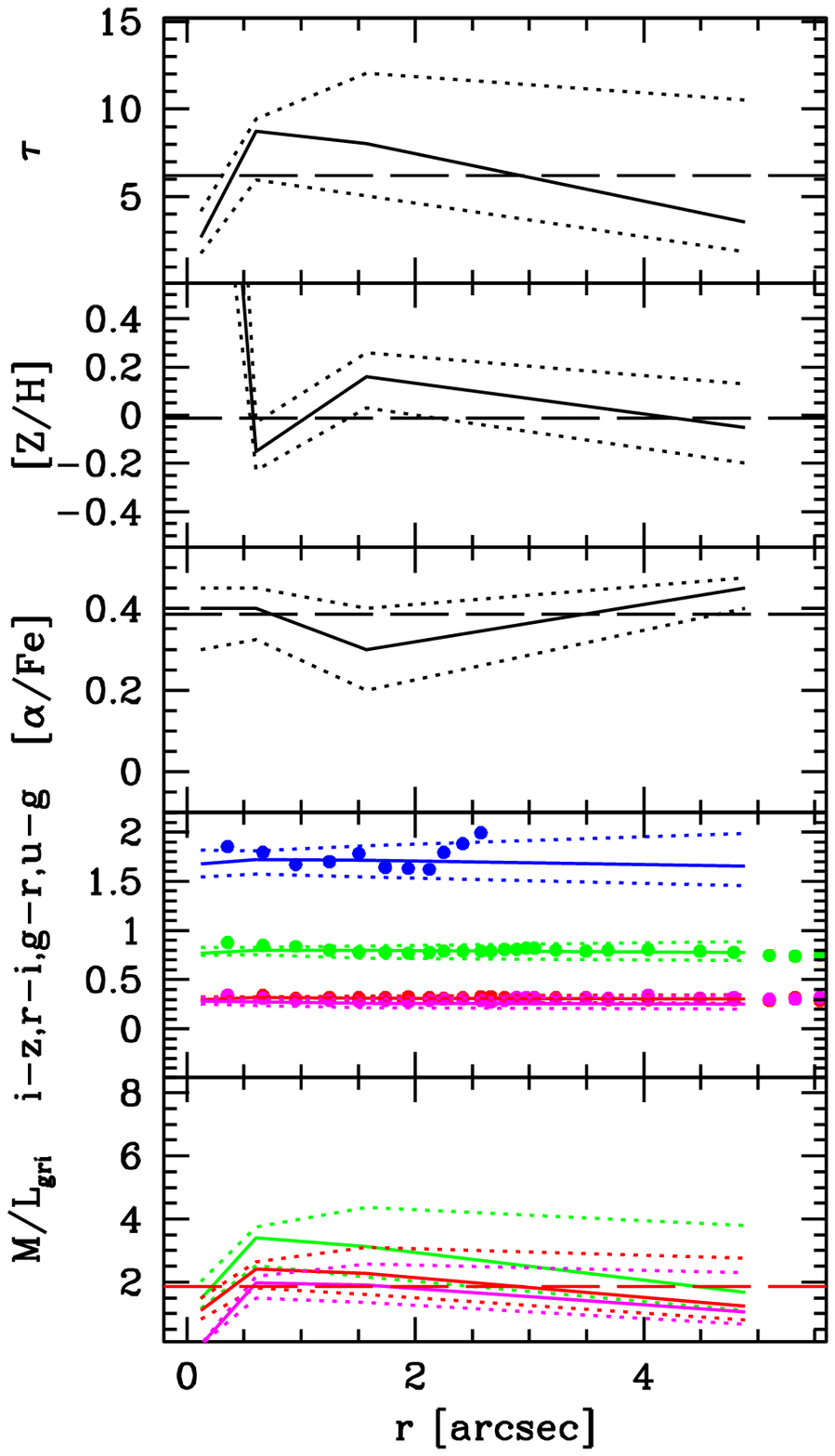}

  \includegraphics[width=9.9cm]{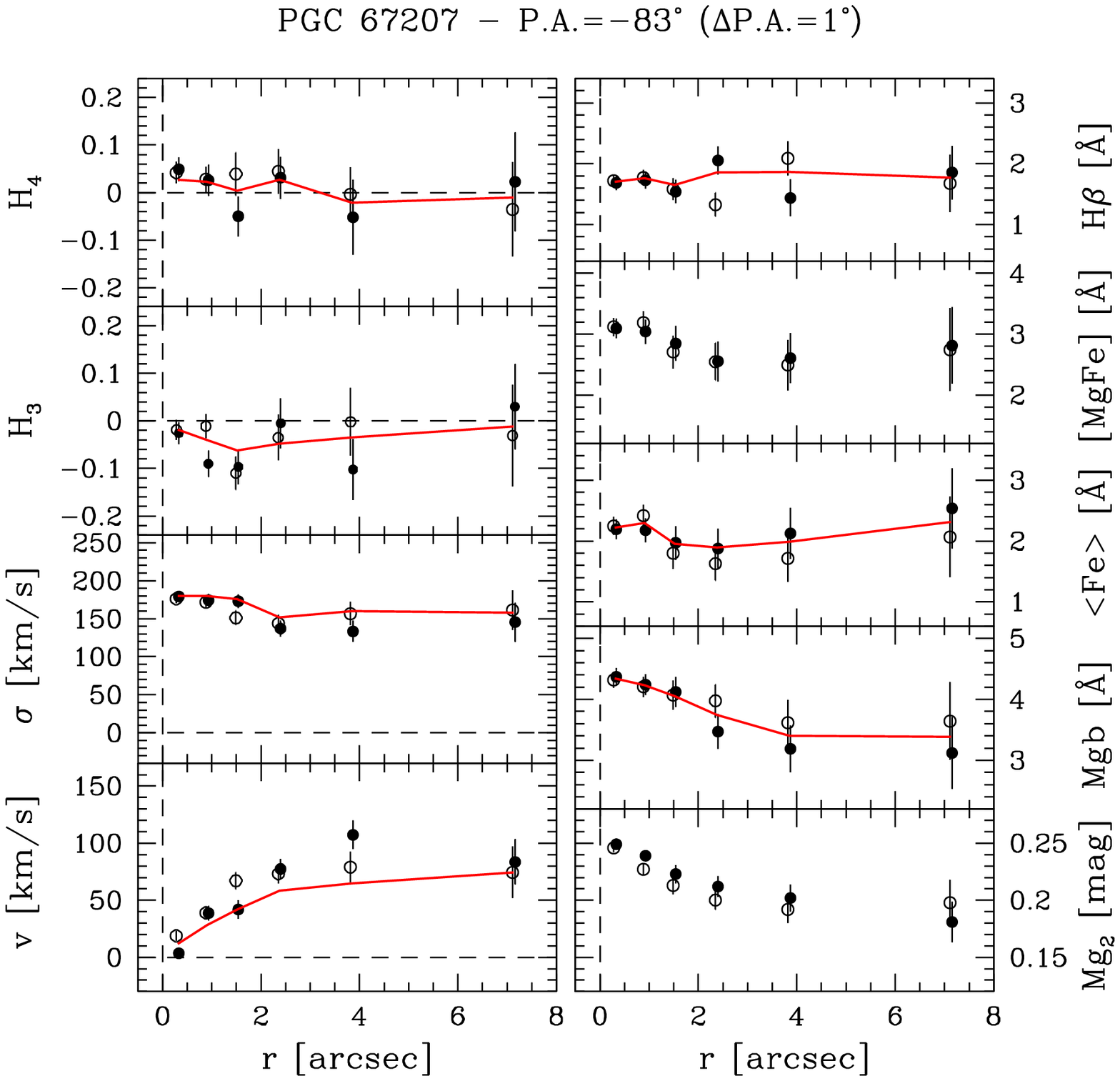}
  \includegraphics[width=5.7cm]{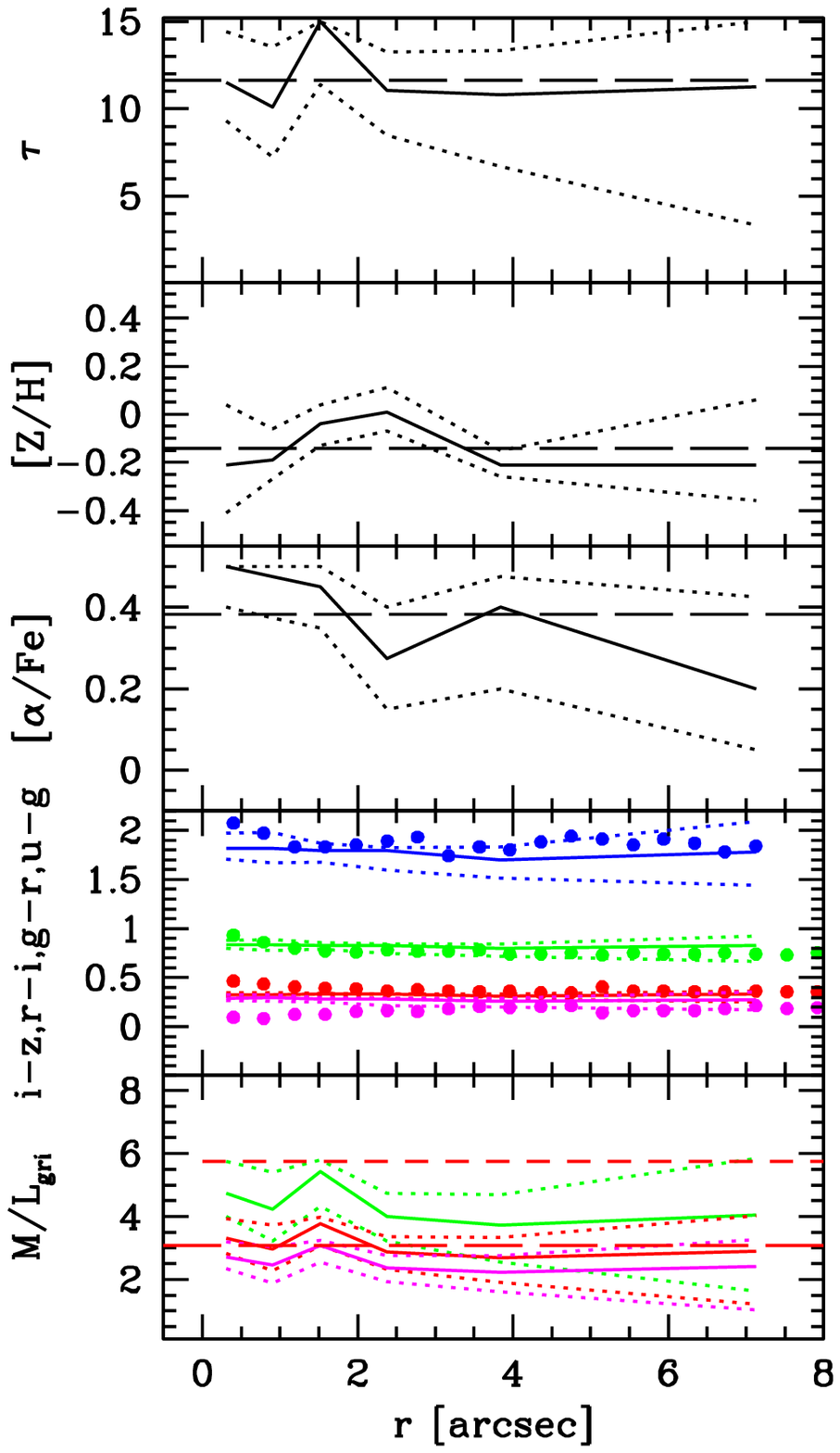}
\caption{{\em continued}}
\end{figure*}

\addtocounter{figure}{-1}
\begin{figure*}
  \includegraphics[width=9.9cm]{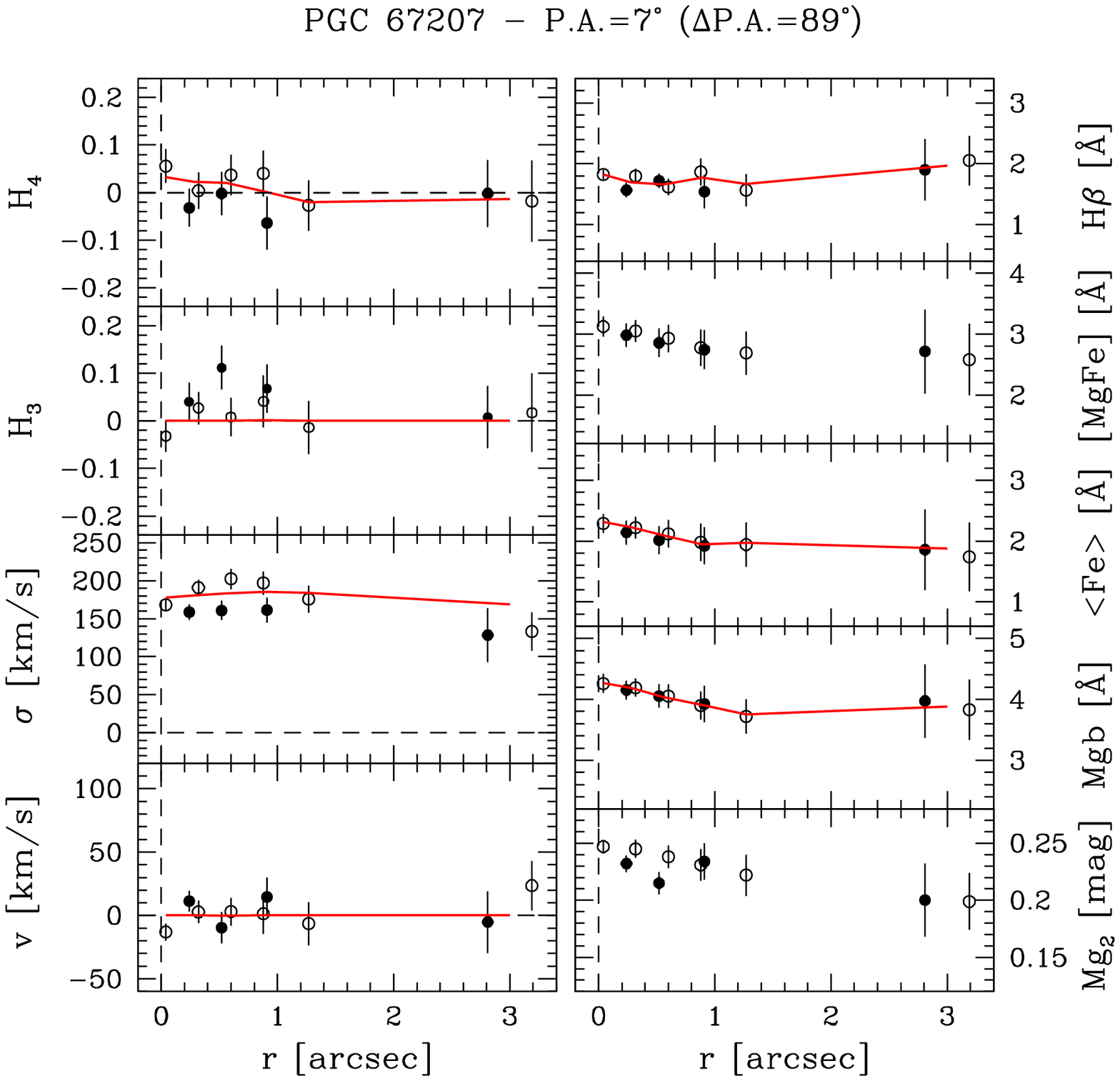}
  \includegraphics[width=5.7cm]{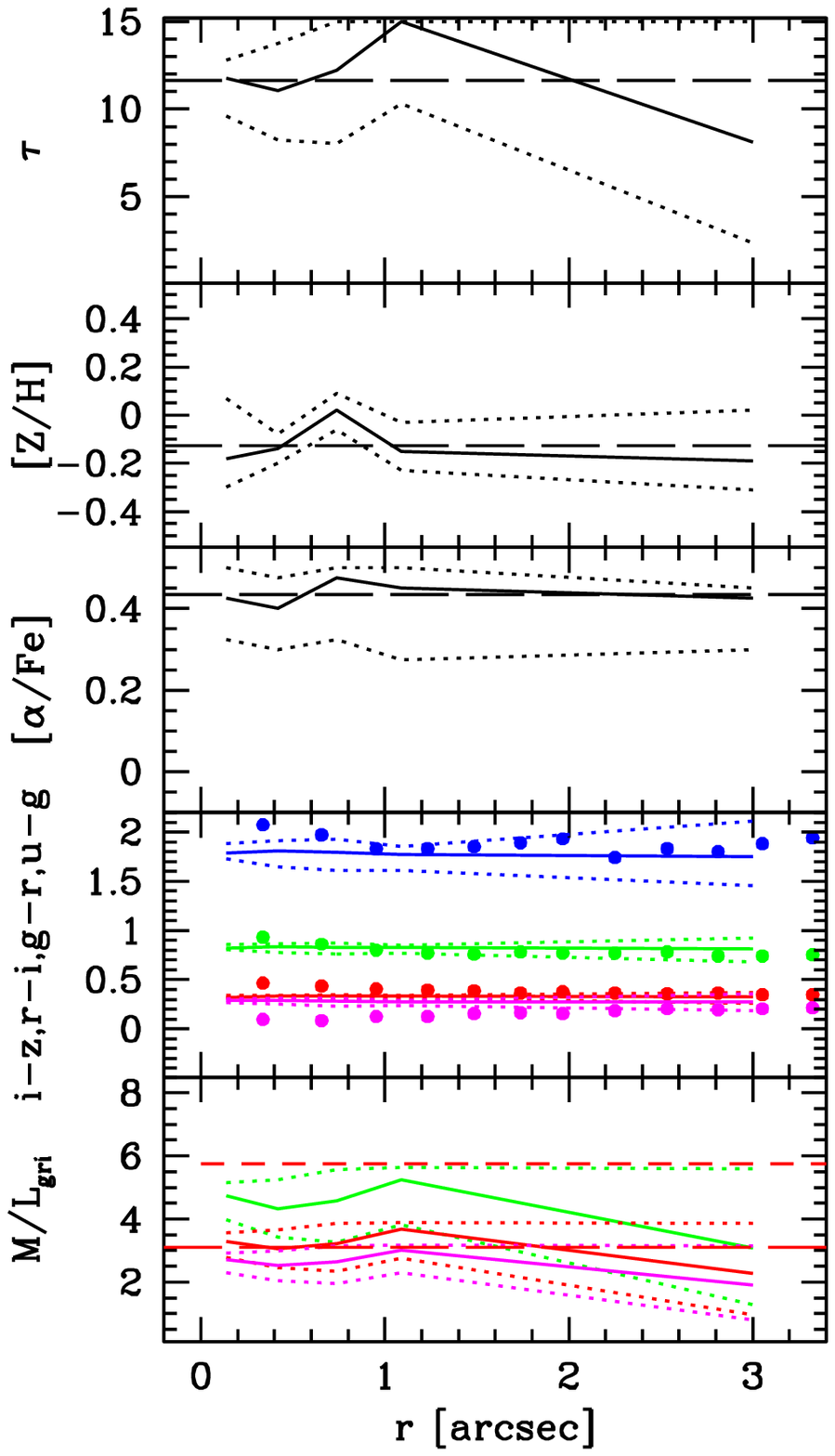}

  \includegraphics[width=9.9cm]{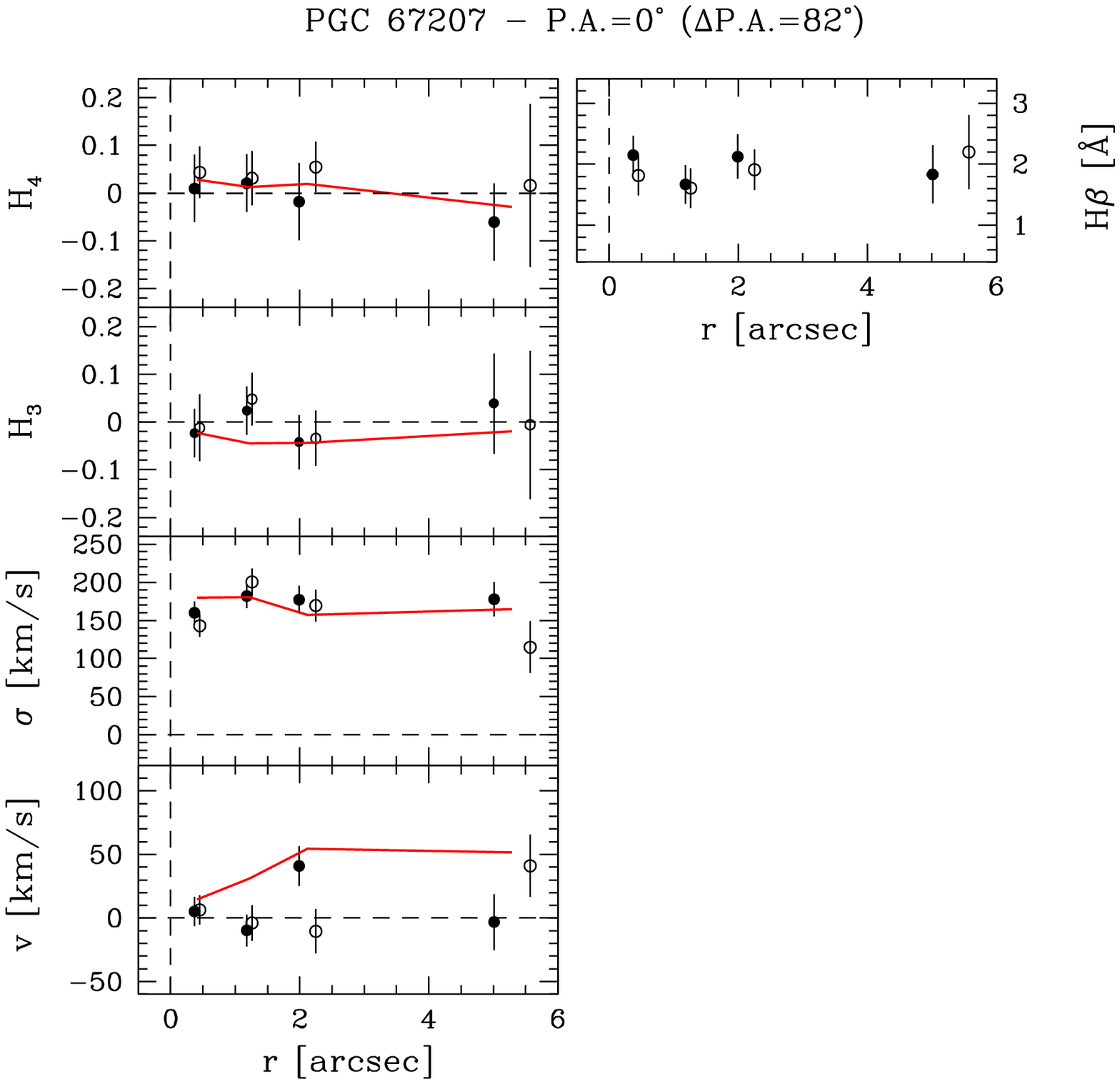}
  \caption{{\em continued}} 
\end{figure*}

\subsection{Line-strength indices}
\label{sec_lineindex}

We measured the Lick/IDS \Hb, Fe$_{5270}$, Fe$_{5335}$, \Mgb,
and \Mgd\ \citep{Faber1985, Worthey1994} and combined
\Fe\ \citep{Gorgas1990} and \MgFe\ \citep{Gonzalez1993} line-strength
indices. We rebinned the spectra in the dispersion direction as well
in the radial one as done for measuring the stellar kinematics. We
took into account for the difference between our spectral resolution
and that of the Lick/IDS system \citep[$\rm FWHM = 8.4$
  \AA;][]{Worthey1997} by degrading our spectra through a Gaussian
convolution to match the Lick/IDS resolution before measuring the
line-strength indices.
We calibrated our measurements to the Lick/IDS system. To this aim, we
calculated the systematic offsets to be applied to our data as the
mean of the differences between the line-strength indices we measured
for the available template G and K giant stars and those obtained by
\citet{Worthey1994}. We applied the same offsets as \citet{Wegner2012}
to data obtained in runs 1 and 2. The wavelength range of run 3
brackets only the \Hb\ line-strength index. We did not apply any
offset, since the \Hb\ values measured at $\pa=0^\circ$ for NGC~7113
in run 3 agree within the errors with those measured in run 1
(Fig.~\ref{fig_kinematics}). The offsets derived in run 4 were
neglected too being smaller than the mean error of the differences.

We derived the errors on the line-strength indices from photon
statistics and CCD read-out noise, and we calibrated them by means of
Monte Carlo simulations. The folded radial profiles of \Hb , \MgFe,
\Fe, \Mgb, and \Mgd\ for all the sample galaxies are listed in
Table~\ref{tab_indices} and plotted in the middle panels of
Fig.~\ref{fig_kinematics}.

\subsection{Properties of the stellar populations}
\label{sec_stellarpops}

We modelled the \Hb , \Fe, and \Mgb\ line-strength indices following
the method described in \citet{Saglia2010a} and \citet{Pu2010}.
We spline interpolated the SSP models for the Kroupa IMF of
\citet{Maraston1998, Maraston2005} on a fine grid in age
($\tau\,\leq\,15$ Gyr on steps of 0.1 Gyr), metallicity
($-2.25\,\leq\,$\ZH$\,\leq\,0.67$ dex on steps of 0.02 dex), and
overabundances ($-0.3\,\leq\,$\aFe$\,\leq\,0.5$ dex on steps of 0.05
dex).
We derived at each radius the best-fitting values of \age , \ZH , and
\aFe\ by minimizing the difference between the values of \Hb ,
\Fe\ and \Mgb\ line-strength indices we measured in the spectra and
those predicted by the SSP models for a given set of age, metallicity,
and overabundance as done in \citet{Wegner2012}. We took into account
the measurement errors and gave the same weight to the line-strength
indices to derive the best-fitting values of the stellar population
properties to the same accuracy. Finally, we computed the values of
the mass-to-light ratio $\mlkroupa$ and colours $u-g$, $g-r$, $r-i$, and
$i-z$ of the corresponding SSP models.

The folded radial profiles of the stellar-population age, metallicity,
overabundance, and mass-to-light ratio together with the predicted and
measured colour gradients are plotted in the right-hand panels of
Fig.~\ref{fig_kinematics}.
The average values of \age , \ZH , \aFe, and $\mlkroupa$ for the sample
galaxies are reported in Table~\ref{tab_galproperties}. They are in
reasonable agreement with the findings of \citet{Wegner2008}.

NGC~7113 is $\sim8$ Gyr old, with metallicity slightly above solar
(\ZH$\sim0.2$ dex), strong overabundance (\aFe$\sim0.3$ dex), and
$\mlkroupa\sim2.5$ \mlsun.
PGC~67207 is $\sim12$ Gyr old, with sub-solar metallicity
(\ZH$\sim-0.1$ dex), strong overabundance (\aFe$\sim0.4$ dex), and
$\mlkroupa\sim3.1$ \mlsun. In both cases, the predicted colour gradients
of the corresponding SSP agree reasonably well with the measured ones,
whereas the mass-to-light ratio of the stellar component is a factor of 2
smaller than the dynamically derived one.
The barred galaxy PGC~1852 has a solar metallicity (\ZH$\sim0$ dex)
and strong overabundance (\aFe$\sim0.4$ dex). Its age (\age$\sim7$
Gyr) and SSP mass-to-light ratio ($\mlkroupa\sim2.0$ \mlsun) are smaller
than those of the other two galaxies, in agreement with the later
morphological type. The older ages derived in the galactic centre
could be related to the bulge component.

\begin{landscape}
\renewcommand{\tabcolsep}{3pt}
\begin{table}  
\caption{The average properties of stellar populations and DM haloes of 
  the sample galaxies. \label{tab_galproperties}}
\begin{center}
\begin{small}
\begin{tabular}{lccrrrcccccccccc}    
\hline 
\noalign{\smallskip}   
\multicolumn{1}{c}{Galaxy} &
\multicolumn{1}{c}{$\reff$} &   
\multicolumn{1}{c}{$\renc$} &    
\multicolumn{1}{c}{$\sigma$} &  
\multicolumn{1}{c}{$\tau$} & 
\multicolumn{1}{c}{\ZH} &   
\multicolumn{1}{c}{\aFe} & 
\multicolumn{1}{c}{$\mlkroupa$} &   
\multicolumn{1}{c}{$\mldyn$} &
\multicolumn{1}{c}{$\log{\avgrhohalo}$} &    
\multicolumn{1}{c}{$\halofrac$} &   
\multicolumn{1}{c}{$\log{\avgrhodmkroupa}$} &    
\multicolumn{1}{c}{$\dmkroupafrac$} \\   
\multicolumn{1}{c}{} &
\multicolumn{1}{c}{(kpc)} &
\multicolumn{1}{c}{(kpc)} &  
\multicolumn{1}{c}{(\kms)} &  
\multicolumn{1}{c}{(Gyr)} & 
\multicolumn{1}{c}{(dex)} &   
\multicolumn{1}{c}{(dex)} & 
\multicolumn{1}{c}{(\mlsun)} &   
\multicolumn{1}{c}{(\mlsun)} & 
\multicolumn{1}{c}{} &   
\multicolumn{1}{c}{} &     
\multicolumn{1}{c}{} &   
\multicolumn{1}{c}{} \\   
\multicolumn{1}{c}{(1)} & 
\multicolumn{1}{c}{(2)} &   
\multicolumn{1}{c}{(3)} & 
\multicolumn{1}{c}{(4)} &   
\multicolumn{1}{c}{(5)} & 
\multicolumn{1}{c}{(6)} & 
\multicolumn{1}{c}{(7)} &
\multicolumn{1}{c}{(8)} & 
\multicolumn{1}{c}{(9)} & 
\multicolumn{1}{c}{(10)} & 
\multicolumn{1}{c}{(11)} & 
\multicolumn{1}{c}{(12)} & 
\multicolumn{1}{c}{(13)} \\ 
\noalign{\smallskip}   
\hline
\noalign{\smallskip}       
NGC~7113  & 8.7 & 17.5 & $178\pm1$ & $ 8.0\pm2.2$ & $ 0.24\pm0.08$ & $0.30\pm0.06$ & $2.52\pm0.50$ & $7.09\pm1.40$ & $-1.89^{+0.39}_{-0.60}$ & $0.36^{+0.40}_{-0.20}$ & $-1.77^{+0.30}_{-0.40}$ & $0.70^{+0.15}_{-0.10}$ \\
PGC~1852  & ... &  ... & $ 95\pm6$ & $ 6.9\pm0.6$ & $-0.04\pm0.06$ & $0.36\pm0.03$ & $1.96\pm0.58$ & ...  & ... & ... & ... & ... \\
PGC~67207 & 4.1 &  8.2 & $164\pm5$ & $11.6\pm0.1$ & $-0.13\pm0.01$ & $0.41\pm0.03$ & $3.05\pm0.05$ & $6.71\pm0.85$ & $-1.56^{+0.37}_{-1.10}$ & $0.11^{+0.17}_{-0.10}$ & $-1.34^{+0.26}_{-0.34}$ &  $0.49^{+0.10}_{-0.06}$ \\
\noalign{\smallskip}       
\hline
\end{tabular} 
\end{small}
\end{center}     
\begin{minipage}{24cm}
{\em Note.} Column (1): name.  Column (2): effective radius.  Column
(3): radius corresponding to $2\reff$ for calculating the DM average
mass density.  Column (4): average velocity dispersion. Column (5):
average age. Column (6): average metallicity. Column (7): average
$\alpha$-elements overabundance.  Column (8): stellar-population
mass-to-light ratio in the $r$ band. Column (9): dynamical
mass-to-light ratio of all the mass following the light in the $r$
band. Column (10): DM-halo average mass density within $\renc$ with
$\rhohalo$ given in \msun pc$^{-3}$.  Column (11): DM-halo
spherically-averaged mass fraction within $\reff$. Column (12): total
DM average mass density within $\renc$ assuming a Kroupa IMF with
$\rhodmkroupa$ given in \msun pc$^{-3}$.  Column (13): total DM
spherically-averaged mass fraction within $\reff$ assuming a Kroupa
IMF.
\end{minipage}
\end{table}  
\end{landscape}  

\section{Dynamical modelling} 
\label{sec_dynamics} 
 
We skipped the barred galaxy PGC~1852 and derived the mass
distribution of NGC~7113 and PGC~67207 as we did for the ETGs in Coma
\citep{Thomas2007, Thomas2011} and Abell~262 \citep{Wegner2012}
clusters. We used the axisymmetric dynamical models by
\citet{Thomas2004, Thomas2005} to obtain the mass-to-light ratio
$\mldyn$ of the mass that follows light and the structural parameters
of the DM halo without assuming any particular orbit structure.

To this aim, we modelled the total mass density distribution as
\begin{equation} 
\label{eq:rho} 
\rho = \rhoast + \rhohalo = \mldyn \nu + \frac{V_{\rm c}^2}{4 \pi G} 
  \frac{3 r_{\rm c}^2+r^2}{(r_{\rm c}^2+r^2)^2} 
\end{equation} 
where $\rhoast$ is the mass density of the (luminous and dark)
matter that is distributed like the stars and $\rhohalo$ is the mass
density of the DM distributed in the halo \citep{Thomas2007,
  Wegner2012}.  We adopted a spherical cored logarithmic DM halo with
asymptotically constant circular velocity $V_{\rm c}$ and constant
mass density inside $r_{\rm c}$ \citep{Binney1987}.
We derived the radial profile of the stellar luminosity density
$\nu$ by deprojecting as in \citet{Magorrian1999} the S\'ersic
surface-brightness profile obtained from the photometric decomposition
and by taking into account the seeing effects as in
\citet{Rusli2011}. We assumed both the galaxies to be seen edge on
($i=90^\circ$). This makes the deprojection unique. We do not expect
that this assumption affects our conclusions about the mass structure
of the modelled galaxies, given our sparse kinematic sampling. For the
same reason we do not discuss the orbital structure in detail.

Fig.~\ref{fig_isophotes} shows the radial profiles of the photometric
parameters of NGC~7113 and PGC~ 67207, which we derived by projecting
the deprojected stellar luminosity density without applying any
correction for seeing convolution.

We computed the gravitational potential from Poisson's equation
  and calculated thousands of orbits in the resulting fixed potential.
  They were superposed to fit the observed LOSVDs with the constraint
  of the luminosity density and using the maximum entropy technique
  \citep{Richstone1988} optimized for the long-slit setup
  \citep[see][for details]{Thomas2004, Thomas2005, Thomas2007}.

The best-fitting mass models with a DM halo are marginally better than
the ones without ($\Delta \chi^2 = 0.8$ for NGC~7113; $\Delta \chi^2 =
2$ for PGC~67207). The corresponding best-fitting kinematics are given
in Fig.~\ref{fig_kinematics}. Moreover, we computed the DM-halo
average mass density $\avgrhohalo$ within $\renc = 2\reff$ and DM-halo
average mass fraction $\halofrac$ inside $\reff$, which we interpreted
as upper limits given the low $\Delta \chi^2$ values.  They are given
in Table~\ref{tab_galproperties} together with the best-fitting
$\mldyn$ and the values of $\reff$ and $\renc$ in physical units. We
opted to average the DM-halo mass density inside the larger $\renc$ to
be less sensitive to slope uncertainties in the radial profile of the
mass density.

Finally, we derived the total DM mass density under the assumption of
Kroupa IMF as
\begin{equation} 
\label{eq:rhodmkroupa} 
\rhodmkroupa = \rhohalo + (\mldyn - \mlkroupa) \times \nu
\end{equation} 
by summing the DM distributed in the halo and that following the
light so closely that it is captured by $\mldyn$ \citep{Thomas2007,
  Wegner2012}.  The corresponding total DM average mass density
$\avgrhodmkroupa$ within $\renc$ and total DM average mass fraction
$\dmkroupafrac$ inside $\reff$ are reported in
Table~\ref{tab_galproperties} too.

\section{Discussion and conclusions}
\label{sec_conclusions}

\begin{figure}
\begin{center}
\includegraphics[width=9cm]{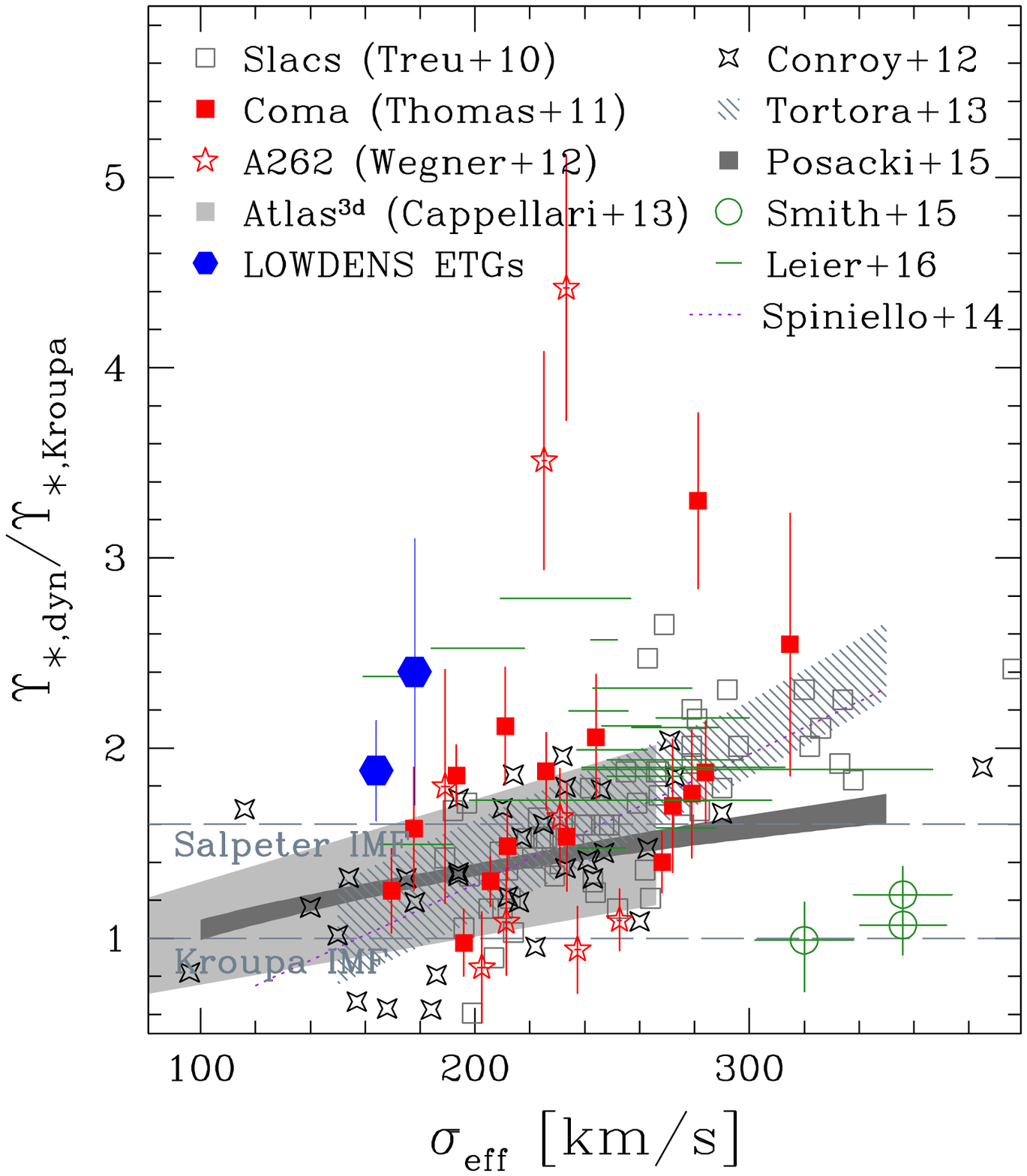}
\end{center}
\caption{The $\ratml$ ratio between the dynamical and
  stellar-population mass-to-light ratios of the ETGs NGC~7113 and
  PGC~67207 (blue filled hexagons) residing in low-density
  environments as a function of the effective velocity dispersion
  $\sigeff$. The values for the ETGs in the SLACS \citep[][open
    squares]{Treu2010} and ATLAS3D \citep[][light grey
    region]{Cappellari2013} surveys, in Coma \citep[][red filled
    squares]{Thomas2011} and Abell 262 \citep[][red open
    stars]{Wegner2012} clusters, and for those studied by
  \citet[][black open stars]{Conroy2012}, \citet[][green open
    circles]{Smith2015a} and \citet[][green solid segments]{Leier2016}
  are shown for comparison. The range of models of \citet[][shaded
    region]{Tortora2013} and the relations by \citet[][dark grey
    region]{Posacki2015} and \citet[][magenta dashed
    line]{Spiniello2014} are also given. The bottom and top horizontal
  dashed lines correspond to the $\ratml$ values for a Kroupa and
  Salpeter IMF, respectively. \label{fig_ratml}}
\end{figure}

Fig.~\ref{fig_ratml} shows the ratio $\ratml$ between the
mass-to-light ratio from dynamical modelling and stellar-population
analysis as a function of the velocity dispersion $\sigeff$ inside
$\reff$. The values we derived for NGC~7113 and PGC~67207 are compared
to those of ETGs in the SLACS \citep{Treu2010} and ATLAS3D
\citep{Cappellari2013} surveys and in the Coma \citep{Thomas2007,
  Thomas2011} and Abell 262 \citep{Wegner2012} clusters.  The results
for the ETGs studied by \citet{Spiniello2014}, \citet{Conroy2012},
\citet{Tortora2013}, \citet{Posacki2015}, \citet{Smith2015a}, and
\citet{Leier2016} are also shown.  In both the galaxies we analysed
the dynamical mass following the light exceeds the Kroupa value by far
($\ratml\ga2$). The trend of increasing $\ratml$ with $\sigeff$ is
well established in ETGs \citep{Thomas2011, Cappellari2012}. It either
points to a systematic variation of the IMF, from Kroupa-like in
low-mass galaxies to Salpeter (or even more bottom-heavy) in most
massive ones \citep{vanDokkum2011, Conroy2012}, or to the presence of
a DM component distributed similar to the stars. Here we report large
$\ratml$ values for two ETGs with a relatively low velocity dispersion
($\sigeff\la170$ \kms), where the majority of the measured galaxies so
far are characterized by $1<\ratml<1.6$ which correspond to the IMF
regime between the Kroupa and Salpeter limits. We do not find any
correlation between $\ratml$ and the ages, metallicities, and
$\alpha$-element overabundance of the stellar populations like in Coma
and Abell~262 ETGs, for which we made the same assumptions on mass
distribution and stellar populations as in NGC~7113 and PGC~67207.

\begin{figure}
\begin{center}
\includegraphics[width=9cm]{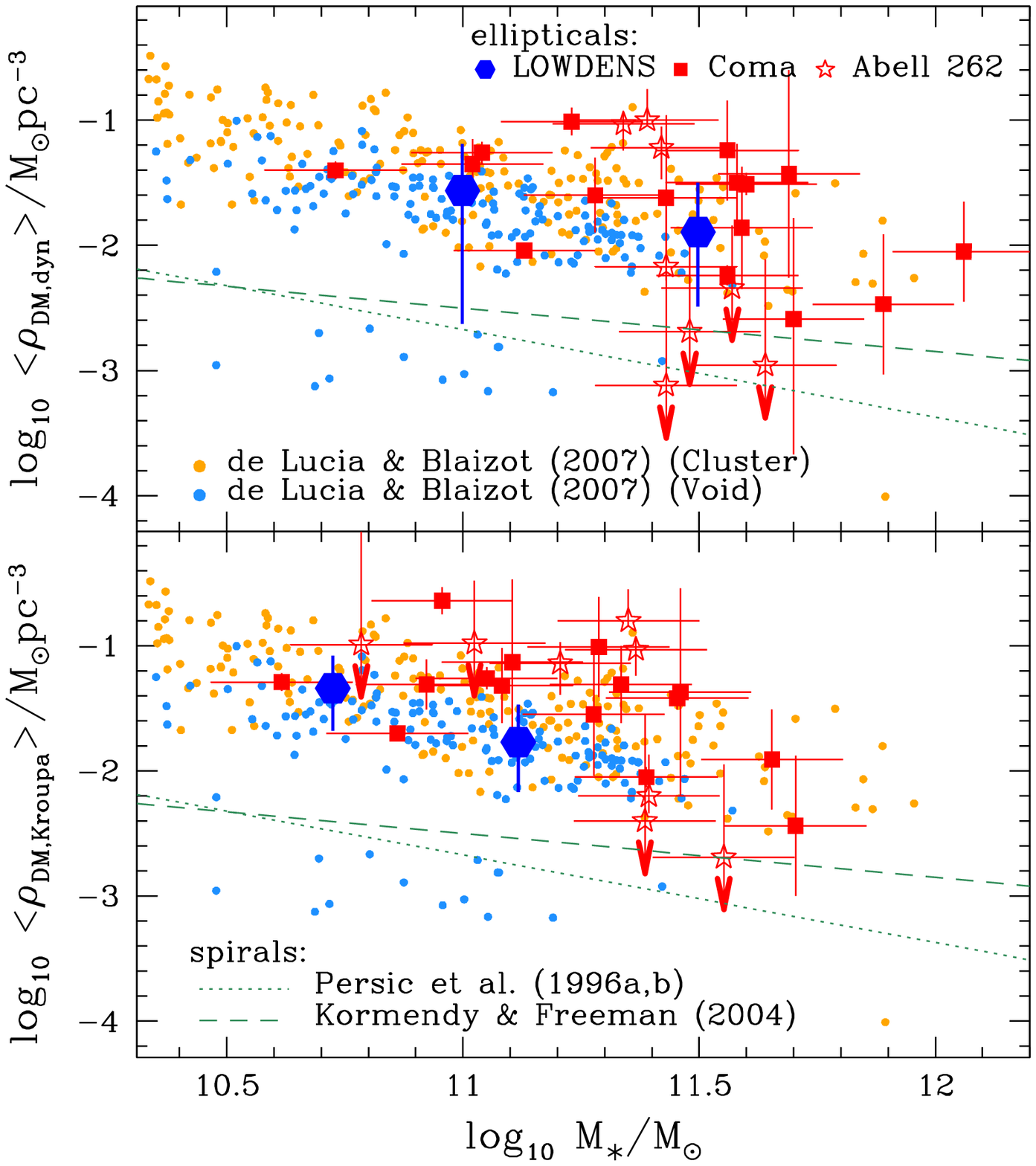}
\end{center}
\caption{DM-halo average mass density (i.e., only halo DM) inside
  $2\reff$ as a function of the mass that follows light ($M_\ast =
  \mldyn \times L_r$, top panel) and total DM average mass density
  (i.e., halo DM + DM following the light) inside $2\reff$ as a
  function of the stellar mass assuming a Kroupa IMF ($M_\ast =
  \mlkroupa \times L_r$, bottom panel) the ETGs NGC~7113 and PGC~67207
  (blue filled hexagons) residing in low-density environments and ETGs
  in Coma \citep[][red filled squares]{Thomas2009a} and Abell 262
  \citep[][red open stars]{Wegner2012} clusters.  Predictions from
  semi-analytic galaxy formation models \citep{Delucia2007} are
  indicated by the small yellow and cyan filled circles for
  high-density regions (galaxy clusters) and low-density regions
  (voids), respectively. The lines show spiral galaxy scaling
  relations from \citet[][green dotted line]{Persic1996a, Persic1996b}
  and \citet[][green dashed line; but see also
  \citealt{Kormendy2016}]{Kormendy2004}.
\label{fig_rhodm}}
\end{figure}

The large $\ratml$ of NGC~7113 and PGC~67207 could be reduced if we
had underestimated the true distances, overestimated the amount of
mass following the light, and/or underestimated $\mlkroupa$ from the
stellar population analysis. Both galaxies follow within $20\%$ the
$r$-band Fundamental Plane of \citet{Hyde2009} and this leads us to
conclude that the assumed distances can not be wrong by much. Our
dynamical models do include a DM halo, so it is not the assumption of
self-consistency that is biasing high our $\mldyn$ estimates
\citep{Thomas2011}. In addition, using the $M_{\rm BH}-\sigma$
relation derived by \citet{Saglia2016} we estimated the mass of the
central supermassive black hole to be $M_{\rm BH}=(0.1-1.4)\times10^8$
\msun\ for $\sigma=170-180$ \kms. The best-fitting models shown in
Fig. \ref{fig_kinematics} predict an enclosed mass $M(R<1\;{\rm
  arcsec}) \sim 7-8\times10^9$ \msun.  So, if the masses of the
supermassive black holes of the two galaxies are not severely
underestimated, we do not expect a significant effect on their
$\mldyn$. But, high-resolution spectroscopy probing the black hole
sphere of influence (expected to be around $0.05$ arcsec) is needed to
fully address this issue. Finally, NGC~7113 has a slightly supersolar
metallicity and is 8 Gyr old while PGC~67297 has slightly subsolar
metallicity and an age of about 12 Gyr. These are rather normal
metallicities and ages for ETGs with similar velocity dispersion and
we reproduced the measured $u-g$, $g-r$, $r-i$, and $i-z$ colours
rather well. This means that we might have underestimated $\mlkroupa$ by
at most 20\% for both galaxies.

Fig. \ref{fig_rhodm} shows the average DM mass inside $2\reff$ as a
function of the stellar mass. Taking the dynamical mass that follows
the light as the stellar mass (i.e., $M_\ast = \mldyn \times L_r$), we
find that the average DM-halo mass density (i.e., only due to halo DM)
inside $2\reff$ of NGC~7113 and PGC~67207 is higher than those of
spiral galaxies \citep{Persic1996a, Persic1996b, Kormendy2004, Kormendy2016}, 
but lower than the ones of cluster ETGs
\citep{Thomas2007, Wegner2012} (Fig. \ref{fig_rhodm}, top panel). This
implies an earlier formation redshift compared to local bright spirals
\citep{Thomas2009a}. Moreover, the DM densities closely match the
predictions of the semi-analytic models for galaxy formation in
low-density regions by \citet{Delucia2007}. This holds in spite of the
degeneracy concerning the interpretation of $\mldyn$, which prevents
discriminating between luminous and DM once they follow close radial
distributions. In fact, the agreement with theoretical predictions is
even better if we assume that stellar mass is minimal by adopting a
Kroupa IMF (i.e., $M_\ast = \mlkroupa \times L_r$) and consider the
total DM average mass density (i.e., due to both the DM in the halo
and that following the light) inside $2\reff$ (Fig. \ref{fig_rhodm},
bottom panel). Without trying to over interpret these results given
the fact the general galaxy population in voids consists of mostly
small late-type spirals and massive ETGs are very rare
\citep{Croton2005, Beygu2016}, the above findings suggest that DM
haloes of ETGs are less dense when these objects live in low-density
regions, even if more extended kinematic data are needed to confirm
this, given the large errors affecting our DM densities. The same
remark applies to the reported DM fractions, which are similar to the
ones derived for cluster ETGs (see Table \ref{tab_galproperties}).

Recent studies of void galaxies \citep{Kreckel2011, Kreckel2012,
  Hoyle2012, Pustilnik2013, Penny2015, Beygu2016} seem to indicate
that the lower density of void regions has a smaller role on setting
the galaxy properties while their nearby environments and DM haloes
are more important. We presented new photometric and spectroscopic
observations of three galaxies living in low-density environments. For
the two of them that are ETGs, our dynamical modelling suggests that
they might have low DM content and large dynamically determined
mass-to-light ratios. Further more extended (and preferably
two-dimensional) stellar kinematics, combined with high-resolution
spectroscopy probing the central regions and NIR measurements of the
strengths of IMF-sensitive lines of these galaxies, are needed to rule
out that it is not the low-density environment of these galaxies that
is causing their peculiar properties.

\section*{Acknowledgments}

We thank the anonymous referee for constructive comments which helped
us to improve this paper. We are grateful to Kathryn Kreckel for
providing us the filtered density contrast of the galaxies from
\citet{Grogin1999} and to Margaret Geller for valuable discussion.
EMC was supported by Padua University through grants 60A02-5857/13,
60A02-5833/14, 60A02-4434/15, CPDA133894, and BIRD164402/16. EMC
acknowledges the Max-Planck-Institut f\"ur extraterrestrische Physik
for the hospitality while this paper was in progress.
This investigation was based on observations obtained at the
Michigan-Dartmouth-MIT (MDM) Observatory and at the Gemini Observatory
under program GN-2011B-9-80. MDM Observatory, Kitt Peak, Arizona is
operated by a consortium of Dartmouth College, the University of
Michigan, Columbia University, the Ohio State University, and Ohio
University.  Gemini Observatory is operated by the Association of
Universities for Research in Astronomy, Inc., under a cooperative
agreement with the NSF on behalf of the Gemini partnership: the
National Science Foundation (United States), the National Research
Council (Canada), CONICYT (Chile), Ministerio de Ciencia,
Tecnolog\'{i}a e Innovaci\'{o}n Productiva (Argentina), and
Minist\'{e}rio da Ci\^{e}ncia, Tecnologia e Inova\c{c}\~{a}o (Brazil).
Parts of the data used in this research were acquired through the Sloan
Digital Sky Survey (SDSS) Archive (http://www.sdss.org/). Funding for
the creation and distribution of the SDSS Archive has been provided by
the Alfred P. Sloan Foundation, the Participating Institutions, the
National Aeronautics and Space Administration, the National Science
Foundation, the US Department of Energy, the Japanese Monbukagakusho,
and the Max Planck Society.  The SDSS is managed by the Astrophysical
Research Consortium (ARC) for the Participating Institutions. The
Participating Institutions are The University of Chicago, Fermilab,
the Institute for Advanced Study, the Japan Participation Group, The
Johns Hopkins University, the Korean Scientist Group, Los Alamos
National Laboratory, the Max-Planck-Institute for Astronomy (MPIA),
the Max-Planck-Institute for Astrophysics (MPA), New Mexico State
University, University of Pittsburgh, University of Portsmouth,
Princeton University, the United States Naval Observatory, and the
University of Washington.
This research also made use of the HyperLeda Database
(http://leda.univ-lyon1.fr/) and NASA/IPAC Extragalactic Database
(NED) which is operated by the Jet Propulsion Laboratory, California
Institute of Technology, under contract with the National Aeronautics
and Space Administration (http://ned.ipac.caltech.edu/).

\bibliographystyle{mn2e}

\appendix
\section{Data tables}

Please note: Oxford University Press is not responsible for the
content or functionality of any supporting materials supplied by the
authors. Any queries (other than missing material) should be directed
to the corresponding author for the article.

\newpage

\renewcommand{\tabcolsep}{3pt}
\begin{table*}  
\caption{Photometric parameters of the sample galaxies from SDSS
  data. \label{tab_photometry_sdss}}
\begin{center}
\begin{small}
\begin{tabular}{rrrrrrcrrrrrrc}      
\hline 
\noalign{\smallskip}   
\multicolumn{1}{c}{$a$} & 
\multicolumn{1}{c}{$\mu_r$} & 
\multicolumn{1}{c}{$e$} & 
\multicolumn{1}{c}{PA} & 
\multicolumn{1}{c}{$\Delta x_{\rm c}$} &
\multicolumn{1}{c}{$\Delta y_{\rm c}$} &
\multicolumn{1}{c}{Err.} &
\multicolumn{1}{c}{$a_3/a$} & 
\multicolumn{1}{c}{$b_3/a$} & 
\multicolumn{1}{c}{$a_4/a$} & 
\multicolumn{1}{c}{$b_4/a$} & 
\multicolumn{1}{c}{$a_6/a$} & 
\multicolumn{1}{c}{$b_6/a$} & 
\multicolumn{1}{c}{Err.} \\
\multicolumn{1}{c}{(arcsec)} & 
\multicolumn{1}{c}{(mag arcsec$^{-2}$)} &
\multicolumn{1}{c}{} &    
\multicolumn{1}{c}{($^\circ$)} & 
\multicolumn{1}{c}{(arcsec)} & 
\multicolumn{1}{c}{(arcsec)} & 
\multicolumn{1}{c}{(arcsec)} & 
\multicolumn{1}{c}{$\times100$} &
\multicolumn{1}{c}{$\times100$} &
\multicolumn{1}{c}{$\times100$} &
\multicolumn{1}{c}{$\times100$} &
\multicolumn{1}{c}{$\times100$} &
\multicolumn{1}{c}{$\times100$} &
\multicolumn{1}{c}{} \\   
\multicolumn{1}{c}{(1)} & 
\multicolumn{1}{c}{(2)} &   
\multicolumn{1}{c}{(3)} & 
\multicolumn{1}{c}{(4)} &   
\multicolumn{1}{c}{(5)} & 
\multicolumn{1}{c}{(6)} &
\multicolumn{1}{c}{(7)} & 
\multicolumn{1}{c}{(8)} &   
\multicolumn{1}{c}{(9)} & 
\multicolumn{1}{c}{(10)} &   
\multicolumn{1}{c}{(11)} & 
\multicolumn{1}{c}{(12)} &
\multicolumn{1}{c}{(13)} & 
\multicolumn{1}{c}{(14)} \\   
\noalign{\smallskip}   
\hline
\noalign{\smallskip}       
\multicolumn{14}{c}{NGC 7113}\\
\noalign{\smallskip}       
\hline
$ 0.207\pm0.049$ & $17.181\pm0.012$ & $0.074\pm0.310$ & $ 83.23\pm129.73$ & $ 0.000$ & $ 0.000$ & $0.014$ & $-0.33$ & $-1.10$ & $-4.24$ & $ 0.57$ & $ 0.79$ & $ 0.28$ & $1.57$\\
$ 0.327\pm0.083$ & $17.230\pm0.013$ & $0.082\pm0.330$ & $ 83.07\pm125.43$ & $-0.008$ & $-0.008$ & $0.023$ & $-0.39$ & $-1.39$ & $-4.91$ & $ 0.60$ & $ 0.38$ & $ 0.25$ & $1.63$\\
$ 0.437\pm0.023$ & $17.214\pm0.013$ & $0.073\pm0.069$ & $ 84.09\pm 29.29$ & $-0.012$ & $-0.008$ & $0.007$ & $-0.52$ & $-0.91$ & $-4.77$ & $ 0.89$ & $ 0.78$ & $-0.19$ & $0.80$\\
$ 0.500\pm0.011$ & $17.415\pm0.014$ & $0.050\pm0.029$ & $ 80.99\pm 17.56$ & $-0.012$ & $-0.008$ & $0.003$ & $-0.15$ & $-0.43$ & $-2.40$ & $-0.04$ & $ 0.26$ & $ 0.11$ & $0.38$\\
$ 0.557\pm0.008$ & $17.465\pm0.014$ & $0.048\pm0.019$ & $ 79.59\pm 11.96$ & $-0.012$ & $-0.008$ & $0.002$ & $-0.06$ & $-0.18$ & $-1.75$ & $-0.19$ & $ 0.16$ & $ 0.19$ & $0.24$\\
\hline
\end{tabular} 
\end{small}
\end{center}     
\begin{minipage}{17cm}
{\em Note.}  A machine-readable version of the full table is available
online.  A few rows of the table are given for showing purpose.
Column (1): semimajor axis. Column (2): azimuthally averaged $r$-band
surface brightness. Column (3): ellipticity defined as $e=1-b/a$ with
$b$ semiminor axis. Column (4): major-axis position angle measured
North through East. Column (5): $x$-coordinate of the centre. Column
(6): $y$-coordinate of the centre.  Column (7): Error on the centre
coordinates defined as Err$={\rm rms_{fit}}/\sqrt{N}$ and derived from
the residual standard deviation ${\rm rms_{fit}}$ of the ellipse fit
to the $N\leq128$ points of the isophote.  Column (8): third-order
sine Fourier coefficient. Column (9): third-order cosine Fourier
coefficient. Column (10): fourth-order sine Fourier
coefficient. Column (11): fourth-order cosine Fourier
coefficient. Column (12): sixth-order sine Fourier coefficient. Column
(13): sixth-order cosine Fourier coefficient. Column (14): error of
Fourier coefficients defined as
Err$=\sqrt{\frac{\sum_{i=10}^{N/2}\,\left(a_i^2+b_i^2\right)}{N/2-10}}\times\frac{100}{a}$.
\end{minipage}  
\end{table*}    

\clearpage

\renewcommand{\tabcolsep}{3pt}
\begin{table}  
\caption{Photometric parameters of the sample galaxies from Gemini
  data. \label{tab_photometry_gemini}}
\begin{center}
\begin{small}
\begin{tabular}{rrrr}      
\hline 
\noalign{\smallskip}
\multicolumn{2}{c}{NGC 7113} & 
\multicolumn{2}{c}{PGC 67207} \\
\multicolumn{1}{c}{$a$} & 
\multicolumn{1}{c}{$\mu_r$} & 
\multicolumn{1}{c}{$a$} & 
\multicolumn{1}{c}{$\mu_r$} \\
\multicolumn{1}{c}{(arcsec)} & 
\multicolumn{1}{c}{(mag arcsec$^{-2}$)} &
\multicolumn{1}{c}{(arcsec)} & 
\multicolumn{1}{c}{(mag arcsec$^{-2}$)} \\
\multicolumn{1}{c}{(1)} & 
\multicolumn{1}{c}{(2)} &   
\multicolumn{1}{c}{(3)} & 
\multicolumn{1}{c}{(4)} \\   
\noalign{\smallskip}   
\hline
\noalign{\smallskip}       
$0.007\pm0.001$ & $ 15.915\pm0.010$ & $0.007\pm0.001$ & $15.204\pm0.010$\\ 
$0.080\pm0.001$ & $ 16.043\pm0.004$ & $0.030\pm0.001$ & $15.329\pm0.009$\\ 
$0.152\pm0.001$ & $ 16.195\pm0.003$ & $0.086\pm0.001$ & $15.490\pm0.008$\\ 
$0.184\pm0.001$ & $ 16.320\pm0.002$ & $0.100\pm0.001$ & $15.619\pm0.007$\\ 
$0.216\pm0.001$ & $ 16.420\pm0.002$ & $0.116\pm0.002$ & $15.709\pm0.007$\\ 
$0.248\pm0.001$ & $ 16.519\pm0.002$ & $0.132\pm0.002$ & $15.808\pm0.007$\\ 
$0.283\pm0.001$ & $ 16.621\pm0.002$ & $0.148\pm0.001$ & $15.920\pm0.007$\\ 
$0.321\pm0.001$ & $ 16.723\pm0.002$ & $0.165\pm0.001$ & $16.025\pm0.005$\\ 
$0.359\pm0.001$ & $ 16.824\pm0.002$ & $0.186\pm0.002$ & $16.126\pm0.005$\\ 
$0.401\pm0.001$ & $ 16.924\pm0.001$ & $0.211\pm0.002$ & $16.225\pm0.005$\\ 
$0.443\pm0.001$ & $ 17.025\pm0.001$ & $0.240\pm0.002$ & $16.333\pm0.005$\\ 
$0.488\pm0.001$ & $ 17.127\pm0.001$ & $0.270\pm0.002$ & $16.437\pm0.005$\\ 
$0.540\pm0.001$ & $ 17.227\pm0.001$ & $0.302\pm0.002$ & $16.540\pm0.004$\\ 
$0.595\pm0.001$ & $ 17.328\pm0.001$ & $0.335\pm0.002$ & $16.643\pm0.003$\\ 
$0.649\pm0.001$ & $ 17.431\pm0.001$ & $0.380\pm0.002$ & $16.749\pm0.003$\\ 
$0.713\pm0.002$ & $ 17.530\pm0.001$ & $0.425\pm0.002$ & $16.854\pm0.003$\\ 
$0.776\pm0.002$ & $ 17.631\pm0.001$ & $0.468\pm0.002$ & $16.960\pm0.003$\\ 
$0.845\pm0.002$ & $ 17.733\pm0.001$ & $0.516\pm0.002$ & $17.068\pm0.002$\\ 
$0.915\pm0.002$ & $ 17.834\pm0.001$ & $0.563\pm0.002$ & $17.175\pm0.002$\\ 
$1.000\pm0.001$ & $ 17.939\pm0.001$ & $0.622\pm0.002$ & $17.282\pm0.002$\\ 
$1.088\pm0.001$ & $ 18.043\pm0.001$ & $0.679\pm0.002$ & $17.389\pm0.002$\\ 
$1.181\pm0.002$ & $ 18.145\pm0.001$ & $0.741\pm0.003$ & $17.500\pm0.002$\\ 
$1.281\pm0.002$ & $ 18.249\pm0.001$ & $0.807\pm0.003$ & $17.611\pm0.002$\\ 
$1.392\pm0.003$ & $ 18.352\pm0.001$ & $0.877\pm0.004$ & $17.724\pm0.002$\\ 
$1.522\pm0.003$ & $ 18.455\pm0.001$ & $0.959\pm0.004$ & $17.840\pm0.002$\\ 
$1.648\pm0.003$ & $ 18.560\pm0.001$ & $1.044\pm0.004$ & $17.959\pm0.002$\\ 
$1.795\pm0.004$ & $ 18.665\pm0.001$ & $1.137\pm0.004$ & $18.075\pm0.002$\\ 
$1.958\pm0.004$ & $ 18.770\pm0.001$ & $1.240\pm0.004$ & $18.196\pm0.002$\\ 
$2.144\pm0.006$ & $ 18.876\pm0.001$ & $1.356\pm0.005$ & $18.319\pm0.002$\\ 
$2.348\pm0.007$ & $ 18.980\pm0.001$ & $1.484\pm0.005$ & $18.442\pm0.001$\\ 
$2.576\pm0.007$ & $ 19.086\pm0.001$ & $1.631\pm0.007$ & $18.571\pm0.001$\\ 
$2.795\pm0.008$ & $ 19.192\pm0.001$ & $1.792\pm0.008$ & $18.704\pm0.001$\\ 
$3.032\pm0.009$ & $ 19.298\pm0.001$ & $1.974\pm0.010$ & $18.840\pm0.001$\\ 
$3.276\pm0.010$ & $ 19.410\pm0.001$ & $2.213\pm0.012$ & $18.979\pm0.001$\\ 
$3.535\pm0.013$ & $ 19.520\pm0.001$ & $2.469\pm0.014$ & $19.123\pm0.001$\\ 
$3.806\pm0.016$ & $ 19.630\pm0.001$ & $2.752\pm0.016$ & $19.272\pm0.001$\\ 
$4.084\pm0.016$ & $ 19.741\pm0.001$ & $3.069\pm0.018$ & $19.430\pm0.001$\\ 
$4.381\pm0.018$ & $ 19.853\pm0.001$ & $3.394\pm0.022$ & $19.592\pm0.001$\\ 
$4.709\pm0.021$ & $ 19.971\pm0.001$ & $3.766\pm0.023$ & $19.758\pm0.001$\\ 
$5.056\pm0.025$ & $ 20.088\pm0.001$ & $...          $ & $...           $\\
$5.457\pm0.027$ & $ 20.208\pm0.001$ & $...          $ & $...           $\\
$5.859\pm0.031$ & $ 20.326\pm0.001$ & $...          $ & $...           $\\
$6.335\pm0.031$ & $ 20.445\pm0.001$ & $...          $ & $...           $\\
$6.829\pm0.038$ & $ 20.572\pm0.001$ & $...          $ & $...           $\\
\noalign{\smallskip}       
\hline
\end{tabular} 
\end{small}
\end{center}     
\begin{minipage}{8.5cm}
{\em Note.} Columns (1) and (3): semimajor axis. Columns (2) and (4):
azimuthally averaged $r$-band surface brightness.
\end{minipage}
\end{table}   

\clearpage

\renewcommand{\tabcolsep}{3pt}
\begin{table*}  
\caption{Stellar kinematics of the sample galaxies. 
  \label{tab_kinematics}}     
\begin{center}
\begin{small}
\begin{tabular}{rrrrrrc}	 			                        
\hline 
\noalign{\smallskip}
\multicolumn{1}{c}{$r$} & 
\multicolumn{1}{c}{$V$} & 
\multicolumn{1}{c}{$\sigma$} & 
\multicolumn{1}{c}{$H_3$} &
\multicolumn{1}{c}{$H_4$} &
\multicolumn{1}{c}{PA} &
\multicolumn{1}{c}{Run} \\
\multicolumn{1}{c}{(arcsec)} & 
\multicolumn{1}{c}{(\kms)} &
\multicolumn{1}{c}{(\kms)} & 
\multicolumn{1}{c}{} &
\multicolumn{1}{c}{} &
\multicolumn{1}{c}{($^\circ$)} &
\multicolumn{1}{c}{} \\
\multicolumn{1}{c}{(1)} & 
\multicolumn{1}{c}{(2)} &   
\multicolumn{1}{c}{(3)} & 
\multicolumn{1}{c}{(4)} & 
\multicolumn{1}{c}{(5)} &   
\multicolumn{1}{c}{(6)} & 
\multicolumn{1}{c}{(7)} \\
\noalign{\smallskip}   
\hline
\noalign{\smallskip}       
\multicolumn{7}{c}{NGC 7113}\\
\noalign{\smallskip}   
\hline
\noalign{\smallskip}       
$ -7.86$ & $  32.4\pm14.6$ & $165.6\pm18.8$ & $ 0.035\pm0.070$ & $ 0.049\pm0.064$ & $  0$ & 1\\
$ -4.19$ & $  27.9\pm10.0$ & $155.6\pm12.9$ & $ 0.011\pm0.056$ & $ 0.057\pm0.045$ & $  0$ & 1\\
$ -2.72$ & $   5.6\pm 7.8$ & $155.4\pm10.2$ & $ 0.041\pm0.038$ & $ 0.023\pm0.048$ & $  0$ & 1\\
$ -1.85$ & $   1.2\pm 8.5$ & $179.8\pm10.9$ & $-0.063\pm0.036$ & $ 0.037\pm0.043$ & $  0$ & 1\\
$ -1.24$ & $ -15.2\pm 6.8$ & $170.8\pm 9.5$ & $ 0.023\pm0.034$ & $ 0.085\pm0.034$ & $  0$ & 1\\
$ -0.64$ & $   8.8\pm 5.3$ & $182.9\pm 6.3$ & $-0.027\pm0.025$ & $-0.009\pm0.027$ & $  0$ & 1\\
$ -0.03$ & $   2.4\pm 4.6$ & $188.9\pm 5.8$ & $-0.037\pm0.021$ & $ 0.011\pm0.024$ & $  0$ & 1\\
$  0.58$ & $   1.1\pm 4.7$ & $184.5\pm 7.0$ & $-0.032\pm0.022$ & $ 0.104\pm0.025$ & $  0$ & 1\\
$  1.18$ & $  -2.2\pm 6.1$ & $196.7\pm 8.2$ & $ 0.017\pm0.027$ & $ 0.075\pm0.028$ & $  0$ & 1\\
$  1.79$ & $ -24.5\pm 7.6$ & $191.4\pm10.0$ & $ 0.016\pm0.035$ & $ 0.055\pm0.034$ & $  0$ & 1\\
$  2.66$ & $ -13.5\pm 8.8$ & $193.1\pm11.7$ & $ 0.064\pm0.037$ & $ 0.071\pm0.040$ & $  0$ & 1\\
$  4.13$ & $  -0.0\pm10.7$ & $189.0\pm13.7$ & $ 0.028\pm0.048$ & $ 0.053\pm0.043$ & $  0$ & 1\\
$  7.16$ & $ -23.9\pm14.5$ & $147.9\pm16.3$ & $ 0.015\pm0.080$ & $-0.031\pm0.068$ & $  0$ & 1\\
$ -7.65$ & $   4.1\pm20.5$ & $195.1\pm19.6$ & $ 0.038\pm0.081$ & $-0.039\pm0.068$ & $  0$ & 3\\
$ -2.90$ & $ -18.3\pm13.9$ & $168.0\pm18.4$ & $-0.048\pm0.068$ & $ 0.076\pm0.062$ & $  0$ & 3\\
$ -1.74$ & $  -7.5\pm14.2$ & $206.2\pm16.3$ & $-0.006\pm0.060$ & $ 0.020\pm0.048$ & $  0$ & 3\\ 
$ -0.93$ & $ -11.4\pm15.9$ & $173.0\pm10.9$ & $-0.009\pm0.054$ & $-0.027\pm0.042$ & $  0$ & 3\\
$ -0.32$ & $ -10.1\pm12.2$ & $168.3\pm15.8$ & $ 0.012\pm0.064$ & $ 0.061\pm0.050$ & $  0$ & 3\\
$  0.09$ & $  -1.2\pm12.4$ & $174.7\pm13.1$ & $ 0.013\pm0.055$ & $-0.017\pm0.054$ & $  0$ & 3\\
$  0.50$ & $  13.1\pm14.6$ & $165.6\pm14.8$ & $-0.033\pm0.049$ & $ 0.013\pm0.067$ & $  0$ & 3\\
$  1.10$ & $  11.1\pm10.6$ & $179.3\pm11.6$ & $-0.038\pm0.045$ & $-0.031\pm0.052$ & $  0$ & 3\\
$  1.92$ & $  11.0\pm13.6$ & $175.4\pm17.0$ & $ 0.030\pm0.055$ & $ 0.037\pm0.068$ & $  0$ & 3\\
$  3.08$ & $ -16.2\pm15.5$ & $180.6\pm19.0$ & $-0.019\pm0.074$ & $ 0.031\pm0.058$ & $  0$ & 3\\
$  7.69$ & $  25.5\pm24.1$ & $188.3\pm29.7$ & $-0.029\pm0.069$ & $ 0.018\pm0.119$ & $  0$ & 3\\
$ -2.84$ & $  34.9\pm18.2$ & $172.4\pm20.3$ & $ 0.036\pm0.078$ & $-0.021\pm0.082$ & $ 90$ & 5\\
$ -0.91$ & $  -4.1\pm 8.3$ & $180.3\pm11.2$ & $-0.016\pm0.054$ & $ 0.056\pm0.046$ & $ 90$ & 5\\
$ -0.51$ & $   8.4\pm10.0$ & $170.7\pm13.5$ & $-0.028\pm0.048$ & $ 0.041\pm0.048$ & $ 90$ & 5\\
$ -0.23$ & $  -3.6\pm 7.5$ & $171.8\pm 9.4$ & $-0.010\pm0.036$ & $ 0.034\pm0.036$ & $ 90$ & 5\\
$  0.05$ & $   0.1\pm 5.5$ & $193.7\pm 6.5$ & $ 0.023\pm0.023$ & $ 0.000\pm0.027$ & $ 90$ & 5\\
$  0.54$ & $  -1.8\pm13.2$ & $183.9\pm14.5$ & $-0.007\pm0.045$ & $ 0.065\pm0.047$ & $ 90$ & 5\\
$  2.89$ & $ -33.8\pm19.4$ & $170.5\pm20.1$ & $-0.033\pm0.076$ & $-0.031\pm0.074$ & $ 90$ & 5\\
\noalign{\smallskip}   
\hline
\noalign{\smallskip}       
\multicolumn{7}{c}{PGC 1852}\\
\noalign{\smallskip}   
\hline
\noalign{\smallskip}       
$-12.14$ & $-182.4\pm22.3$ & $ 41.4\pm24.1$ & $-0.002\pm0.193$ & $ 0.011\pm0.180$ & $-42$ & 1\\
$ -8.95$ & $-194.9\pm 9.6$ & $ 50.4\pm29.9$ & $ 0.039\pm0.121$ & $-0.011\pm0.167$ & $-42$ & 1\\
$ -6.86$ & $-154.5\pm 9.1$ & $ 79.4\pm10.8$ & $-0.003\pm0.094$ & $ 0.056\pm0.059$ & $-42$ & 1\\
$ -5.38$ & $-142.8\pm 7.6$ & $ 87.2\pm12.1$ & $ 0.069\pm0.053$ & $ 0.017\pm0.101$ & $-42$ & 1\\
$ -4.17$ & $ -98.6\pm 8.6$ & $ 72.3\pm 9.7$ & $ 0.001\pm0.096$ & $ 0.020\pm0.059$ & $-42$ & 1\\
$ -2.94$ & $ -85.8\pm 6.2$ & $ 98.9\pm 9.0$ & $ 0.070\pm0.044$ & $ 0.036\pm0.063$ & $-42$ & 1\\
$ -2.06$ & $ -58.9\pm 9.5$ & $109.3\pm 8.2$ & $ 0.078\pm0.046$ & $-0.034\pm0.059$ & $-42$ & 1\\
$ -1.45$ & $ -46.2\pm 6.4$ & $121.8\pm 8.9$ & $ 0.019\pm0.043$ & $ 0.056\pm0.042$ & $-42$ & 1\\
$ -0.85$ & $ -26.3\pm 5.3$ & $125.6\pm 7.0$ & $ 0.010\pm0.036$ & $ 0.018\pm0.035$ & $-42$ & 1\\
$ -0.24$ & $ -16.9\pm 5.0$ & $124.6\pm 5.5$ & $ 0.046\pm0.034$ & $-0.058\pm0.034$ & $-42$ & 1\\
$  0.37$ & $  14.4\pm 4.7$ & $120.8\pm 6.1$ & $-0.010\pm0.031$ & $ 0.019\pm0.034$ & $-42$ & 1\\
$  0.97$ & $  21.8\pm 5.5$ & $127.2\pm 7.0$ & $-0.016\pm0.035$ & $ 0.010\pm0.037$ & $-42$ & 1\\
$  1.58$ & $  46.9\pm 6.3$ & $113.5\pm 8.0$ & $-0.040\pm0.039$ & $ 0.037\pm0.047$ & $-42$ & 1\\
$  2.18$ & $  45.8\pm 7.1$ & $106.3\pm 8.7$ & $-0.031\pm0.048$ & $ 0.022\pm0.052$ & $-42$ & 1\\
$  3.06$ & $  83.3\pm 6.5$ & $ 96.3\pm 8.4$ & $ 0.008\pm0.048$ & $-0.001\pm0.062$ & $-42$ & 1\\
$  4.29$ & $ 128.9\pm 7.8$ & $ 84.0\pm11.7$ & $-0.053\pm0.058$ & $ 0.036\pm0.099$ & $-42$ & 1\\
$  5.51$ & $ 134.3\pm 9.5$ & $ 93.4\pm12.6$ & $-0.042\pm0.077$ & $ 0.059\pm0.074$ & $-42$ & 1\\
$  6.99$ & $ 165.0\pm 8.1$ & $ 91.4\pm12.2$ & $-0.003\pm0.055$ & $-0.012\pm0.100$ & $-42$ & 1\\
$  9.06$ & $ 177.3\pm17.1$ & $ 38.0\pm29.5$ & $-0.013\pm0.126$ & $ 0.002\pm0.152$ & $-42$ & 1\\
$ 12.20$ & $ 189.7\pm34.1$ & $ 37.9\pm25.3$ & $ 0.001\pm0.160$ & $ 0.005\pm0.217$ & $-42$ & 1\\
$ -5.25$ & $  -1.4\pm23.0$ & $ 60.7\pm19.0$ & $ 0.034\pm0.127$ & $-0.023\pm0.121$ & $ 48$ & 2\\ 
$ -1.81$ & $  -5.9\pm 5.8$ & $ 98.7\pm 6.9$ & $-0.001\pm0.049$ & $-0.028\pm0.046$ & $ 48$ & 2\\
$ -0.73$ & $   3.2\pm 5.2$ & $114.9\pm 7.1$ & $ 0.062\pm0.034$ & $ 0.002\pm0.044$ & $ 48$ & 2\\
\noalign{\smallskip}       
\hline
\end{tabular} 
\end{small}
\end{center}     
\end{table*}    

\begin{table*}
\contcaption{}
\begin{center}
\begin{small}
\begin{tabular}{rrrrrrc}      
\hline 
\noalign{\smallskip}
$ -0.12$ & $  -4.8\pm 4.7$ & $110.8\pm 6.5$ & $ 0.050\pm0.032$ & $ 0.041\pm0.039$ & $ 48$ & 2\\
$  0.48$ & $ -11.5\pm 5.3$ & $106.1\pm 7.7$ & $ 0.027\pm0.039$ & $ 0.089\pm0.043$ & $ 48$ & 2\\
$  1.34$ & $  12.4\pm 4.8$ & $117.2\pm 5.7$ & $-0.030\pm0.034$ & $-0.030\pm0.035$ & $ 48$ & 2\\
$  4.52$ & $   7.9\pm22.2$ & $101.1\pm15.0$ & $ 0.010\pm0.120$ & $ 0.012\pm0.109$ & $ 48$ & 2\\
\noalign{\smallskip}                                                                       
\hline                                                                                     
\noalign{\smallskip}                                                                       
\multicolumn{7}{c}{PGC 1852}\\
\noalign{\smallskip}   
\hline
\noalign{\smallskip}       
$ -7.16$ & $  83.5\pm19.4$ & $145.3\pm25.2$ & $ 0.030\pm0.089$ & $ 0.023\pm0.103$ & $-83$ & 1\\
$ -3.87$ & $ 107.3\pm12.6$ & $133.5\pm13.8$ & $-0.102\pm0.064$ & $-0.052\pm0.079$ & $-83$ & 1\\
$ -2.40$ & $  77.3\pm 8.7$ & $137.0\pm10.7$ & $-0.005\pm0.053$ & $ 0.031\pm0.044$ & $-83$ & 1\\
$ -1.54$ & $  41.7\pm 8.2$ & $172.9\pm 8.9$ & $-0.097\pm0.037$ & $-0.050\pm0.042$ & $-83$ & 1\\
$ -0.93$ & $  38.5\pm 5.9$ & $174.5\pm 7.7$ & $-0.090\pm0.027$ & $ 0.026\pm0.032$ & $-83$ & 1\\
$ -0.33$ & $   3.6\pm 4.8$ & $179.0\pm 6.4$ & $-0.026\pm0.023$ & $ 0.049\pm0.025$ & $-83$ & 1\\
$  0.28$ & $ -18.9\pm 4.3$ & $176.1\pm 5.7$ & $ 0.019\pm0.021$ & $ 0.042\pm0.023$ & $-83$ & 1\\
$  0.88$ & $ -39.0\pm 5.1$ & $171.9\pm 6.6$ & $ 0.011\pm0.026$ & $ 0.028\pm0.027$ & $-83$ & 1\\
$  1.49$ & $ -67.3\pm 7.3$ & $151.2\pm 9.2$ & $ 0.110\pm0.035$ & $ 0.039\pm0.045$ & $-83$ & 1\\
$  2.35$ & $ -73.3\pm 8.3$ & $143.9\pm11.1$ & $ 0.035\pm0.048$ & $ 0.045\pm0.047$ & $-83$ & 1\\
$  3.82$ & $ -79.2\pm13.4$ & $156.8\pm15.6$ & $ 0.002\pm0.071$ & $-0.004\pm0.057$ & $-83$ & 1\\
$  7.11$ & $ -74.6\pm22.8$ & $161.4\pm26.2$ & $ 0.031\pm0.107$ & $-0.035\pm0.099$ & $-83$ & 1\\
$ -5.57$ & $ -41.1\pm24.3$ & $115.1\pm34.0$ & $ 0.006\pm0.156$ & $ 0.016\pm0.171$ & $  0$ & 3\\
$ -2.25$ & $  10.4\pm17.5$ & $169.3\pm20.7$ & $ 0.034\pm0.058$ & $ 0.055\pm0.053$ & $  0$ & 3\\
$ -1.26$ & $   4.0\pm14.1$ & $200.4\pm17.7$ & $-0.048\pm0.055$ & $ 0.031\pm0.057$ & $  0$ & 3\\
$ -0.45$ & $  -6.3\pm11.5$ & $143.2\pm14.9$ & $ 0.012\pm0.070$ & $ 0.044\pm0.054$ & $  0$ & 3\\
$  0.37$ & $   5.2\pm11.5$ & $159.9\pm15.3$ & $-0.023\pm0.051$ & $ 0.010\pm0.071$ & $  0$ & 3\\
$  1.18$ & $  -9.8\pm12.6$ & $182.0\pm15.8$ & $ 0.024\pm0.051$ & $ 0.021\pm0.061$ & $  0$ & 3\\
$  1.99$ & $  40.9\pm15.5$ & $177.4\pm18.3$ & $-0.042\pm0.057$ & $-0.018\pm0.081$ & $  0$ & 3\\
$  5.01$ & $  -3.2\pm22.0$ & $177.8\pm22.7$ & $ 0.039\pm0.105$ & $-0.061\pm0.081$ & $  0$ & 3\\
$ -2.81$ & $  -5.3\pm24.0$ & $128.6\pm35.0$ & $ 0.008\pm0.065$ & $-0.002\pm0.070$ & $  7$ & 5\\
$ -0.91$ & $  14.6\pm15.3$ & $161.5\pm16.4$ & $ 0.068\pm0.051$ & $-0.064\pm0.056$ & $  7$ & 5\\
$ -0.52$ & $  -9.6\pm12.4$ & $160.9\pm12.6$ & $ 0.112\pm0.046$ & $-0.002\pm0.046$ & $  7$ & 5\\
$ -0.24$ & $  11.2\pm 8.2$ & $158.6\pm10.0$ & $ 0.040\pm0.040$ & $-0.032\pm0.040$ & $  7$ & 5\\
$  0.04$ & $  13.1\pm 6.8$ & $168.2\pm 9.2$ & $ 0.032\pm0.033$ & $ 0.056\pm0.036$ & $  7$ & 5\\
$  0.32$ & $  -2.9\pm 8.4$ & $191.0\pm 9.5$ & $-0.027\pm0.033$ & $ 0.004\pm0.037$ & $  7$ & 5\\
$  0.60$ & $  -3.1\pm10.9$ & $202.3\pm13.4$ & $-0.008\pm0.041$ & $ 0.037\pm0.043$ & $  7$ & 5\\
$  0.88$ & $  -1.3\pm15.9$ & $196.8\pm15.6$ & $-0.041\pm0.055$ & $ 0.040\pm0.048$ & $  7$ & 5\\
$  1.27$ & $   6.5\pm17.0$ & $175.5\pm17.9$ & $ 0.014\pm0.056$ & $-0.027\pm0.053$ & $  7$ & 5\\
$  3.19$ & $ -23.5\pm19.4$ & $133.1\pm25.4$ & $-0.017\pm0.083$ & $-0.018\pm0.086$ & $  7$ & 5\\
\noalign{\smallskip}       
\hline
\end{tabular} 
\end{small}
\end{center}     
\begin{minipage}{10cm}
{\em Note.} Column (1): radius. Column (2): LOS velocity after
subtraction of systemic velocity. Column (3): LOS velocity dispersion.
Column (4): third-order Gauss-Hermite coefficient. Column (5):
fourth-order Gauss-Hermite coefficient. Column (6): slit position
angle measured North through East. Column(7): observing run.
\end{minipage}  
\end{table*}    

\clearpage

\renewcommand{\tabcolsep}{3pt}
\begin{table*}  
\caption{Line-strength indices of the sample galaxies. 
  \label{tab_indices}}     
\begin{center}
\begin{small}
\begin{tabular}{rrrrrrrc}	 			                        
\hline 
\noalign{\smallskip}
\multicolumn{1}{c}{$r$} & 
\multicolumn{1}{c}{\Hb} & 
\multicolumn{1}{c}{\MgFe} & 
\multicolumn{1}{c}{\Fe} &
\multicolumn{1}{c}{\Mgb} &
\multicolumn{1}{c}{\Mgd} &
\multicolumn{1}{c}{PA} &
\multicolumn{1}{c}{Run} \\
\multicolumn{1}{c}{(arcsec)} & 
\multicolumn{1}{c}{(\AA)} &
\multicolumn{1}{c}{(\AA)} & 
\multicolumn{1}{c}{(\AA)} &
\multicolumn{1}{c}{(\AA)} &
\multicolumn{1}{c}{(mag)} &
\multicolumn{1}{c}{($^\circ$)} &
\multicolumn{1}{c}{} \\
\multicolumn{1}{c}{(1)} & 
\multicolumn{1}{c}{(2)} &   
\multicolumn{1}{c}{(3)} & 
\multicolumn{1}{c}{(4)} & 
\multicolumn{1}{c}{(5)} &   
\multicolumn{1}{c}{(6)} &
\multicolumn{1}{c}{(7)} &
\multicolumn{1}{c}{(8)} \\
\noalign{\smallskip}   
\hline
\noalign{\smallskip}       
\multicolumn{8}{c}{NGC 7113}\\
\noalign{\smallskip}   
\hline
\noalign{\smallskip}       
$ -7.86$ & $2.316\pm0.363$ & $2.893\pm0.574$ & $1.953\pm0.539$ & $4.286\pm0.518$ & $0.246\pm0.014$ & $  0$ & 1\\
$ -4.19$ & $2.290\pm0.215$ & $2.989\pm0.343$ & $2.242\pm0.332$ & $3.983\pm0.325$ & $0.263\pm0.009$ & $  0$ & 1\\
$ -2.72$ & $1.552\pm0.187$ & $2.965\pm0.290$ & $2.241\pm0.282$ & $3.922\pm0.274$ & $0.252\pm0.007$ & $  0$ & 1\\
$ -1.85$ & $1.503\pm0.177$ & $3.025\pm0.288$ & $2.096\pm0.273$ & $4.366\pm0.262$ & $0.283\pm0.007$ & $  0$ & 1\\
$ -1.24$ & $1.574\pm0.150$ & $3.138\pm0.237$ & $2.060\pm0.219$ & $4.780\pm0.215$ & $0.293\pm0.006$ & $  0$ & 1\\
$ -0.64$ & $1.444\pm0.123$ & $3.083\pm0.184$ & $2.263\pm0.178$ & $4.198\pm0.173$ & $0.278\pm0.005$ & $  0$ & 1\\
$ -0.03$ & $1.546\pm0.110$ & $3.374\pm0.158$ & $2.471\pm0.152$ & $4.608\pm0.148$ & $0.277\pm0.004$ & $  0$ & 1\\
$  0.58$ & $1.897\pm0.122$ & $3.284\pm0.161$ & $2.330\pm0.153$ & $4.630\pm0.152$ & $0.274\pm0.004$ & $  0$ & 1\\
$  1.18$ & $1.257\pm0.141$ & $3.200\pm0.181$ & $2.283\pm0.170$ & $4.485\pm0.172$ & $0.260\pm0.004$ & $  0$ & 1\\
$  1.79$ & $2.054\pm0.183$ & $3.152\pm0.228$ & $2.233\pm0.214$ & $4.451\pm0.217$ & $0.245\pm0.006$ & $  0$ & 1\\
$  2.66$ & $1.574\pm0.204$ & $3.094\pm0.265$ & $2.107\pm0.244$ & $4.541\pm0.253$ & $0.263\pm0.006$ & $  0$ & 1\\
$  4.13$ & $1.800\pm0.251$ & $3.349\pm0.331$ & $2.504\pm0.316$ & $4.478\pm0.321$ & $0.245\pm0.008$ & $  0$ & 1\\
$  7.16$ & $2.204\pm0.370$ & $2.793\pm0.509$ & $1.817\pm0.464$ & $4.292\pm0.469$ & $0.234\pm0.012$ & $  0$ & 1\\
$ -7.65$ & $1.594\pm0.502$ & ...             & ...             & ...             & ...             & $  0$ & 3\\
$ -2.90$ & $1.466\pm0.318$ & ...             & ...             & ...             & ...             & $  0$ & 3\\
$ -1.74$ & $1.548\pm0.298$ & ...             & ...             & ...             & ...             & $  0$ & 3\\
$ -0.93$ & $1.599\pm0.257$ & ...             & ...             & ...             & ...             & $  0$ & 3\\
$ -0.32$ & $1.849\pm0.319$ & ...             & ...             & ...             & ...             & $  0$ & 3\\
$  0.09$ & $2.038\pm0.311$ & ...             & ...             & ...             & ...             & $  0$ & 3\\
$  0.50$ & $1.568\pm0.311$ & ...             & ...             & ...             & ...             & $  0$ & 3\\
$  1.10$ & $1.497\pm0.255$ & ...             & ...             & ...             & ...             & $  0$ & 3\\
$  1.92$ & $2.245\pm0.319$ & ...             & ...             & ...             & ...             & $  0$ & 3\\
$  3.08$ & $2.023\pm0.333$ & ...             & ...             & ...             & ...             & $  0$ & 3\\
$  7.69$ & $2.249\pm0.514$ & ...             & ...             & ...             & ...             & $  0$ & 3\\
$ -2.84$ & $1.843\pm0.336$ & $3.050\pm0.866$ & $2.234\pm0.873$ & $4.164\pm0.739$ & $0.224\pm0.021$ & $ 90$ & 5\\
$ -0.91$ & $2.238\pm0.136$ & $3.206\pm0.406$ & $2.562\pm0.430$ & $4.012\pm0.343$ & $0.283\pm0.010$ & $ 90$ & 5\\
$ -0.51$ & $2.168\pm0.113$ & $3.144\pm0.411$ & $2.332\pm0.432$ & $4.240\pm0.324$ & $0.287\pm0.009$ & $ 90$ & 5\\
$ -0.23$ & $1.866\pm0.090$ & $3.301\pm0.322$ & $2.377\pm0.337$ & $4.585\pm0.244$ & $0.272\pm0.007$ & $ 90$ & 5\\
$  0.05$ & $1.629\pm0.080$ & $3.394\pm0.208$ & $2.677\pm0.226$ & $4.303\pm0.165$ & $0.270\pm0.005$ & $ 90$ & 5\\
$  0.54$ & $1.977\pm0.120$ & $3.302\pm0.341$ & $2.493\pm0.348$ & $4.373\pm0.295$ & $0.251\pm0.007$ & $ 90$ & 5\\
$  2.89$ & $2.217\pm0.341$ & $3.374\pm0.310$ & $2.777\pm0.278$ & $4.098\pm0.343$ & $0.266\pm0.010$ & $ 90$ & 5\\
\noalign{\smallskip}   
\hline
\noalign{\smallskip}       
\multicolumn{8}{c}{PGC 1852}\\
\noalign{\smallskip}   
\hline
\noalign{\smallskip}       
$-12.14$ & $1.822\pm0.652$ & $2.167\pm0.714$ & $1.619\pm0.675$ & $2.900\pm0.703$ & $0.214\pm0.020$ & $-42$ & 1\\
$ -8.95$ & $2.313\pm0.461$ & $2.721\pm0.509$ & $2.015\pm0.479$ & $3.674\pm0.501$ & $0.214\pm0.014$ & $-42$ & 1\\
$ -6.86$ & $2.753\pm0.353$ & $2.305\pm0.435$ & $1.383\pm0.379$ & $3.839\pm0.396$ & $0.220\pm0.011$ & $-42$ & 1\\
$ -5.38$ & $1.900\pm0.314$ & $2.784\pm0.387$ & $1.845\pm0.352$ & $4.201\pm0.369$ & $0.223\pm0.010$ & $-42$ & 1\\
$ -4.17$ & $2.181\pm0.286$ & $2.846\pm0.336$ & $1.991\pm0.311$ & $4.069\pm0.325$ & $0.231\pm0.009$ & $-42$ & 1\\
$ -2.94$ & $2.366\pm0.219$ & $2.773\pm0.260$ & $2.079\pm0.249$ & $3.699\pm0.251$ & $0.224\pm0.007$ & $-42$ & 1\\
$ -2.06$ & $1.586\pm0.232$ & $2.859\pm0.276$ & $2.367\pm0.269$ & $3.454\pm0.275$ & $0.231\pm0.008$ & $-42$ & 1\\
$ -1.45$ & $2.194\pm0.196$ & $2.987\pm0.230$ & $2.484\pm0.223$ & $3.593\pm0.230$ & $0.219\pm0.007$ & $-42$ & 1\\
$ -0.85$ & $2.094\pm0.157$ & $2.824\pm0.187$ & $2.216\pm0.179$ & $3.597\pm0.186$ & $0.216\pm0.005$ & $-42$ & 1\\
$ -0.24$ & $1.089\pm0.151$ & $2.819\pm0.174$ & $2.126\pm0.162$ & $3.736\pm0.178$ & $0.213\pm0.005$ & $-42$ & 1\\
$  0.37$ & $1.348\pm0.149$ & $2.870\pm0.158$ & $2.292\pm0.148$ & $3.595\pm0.163$ & $0.203\pm0.005$ & $-42$ & 1\\
$  0.97$ & $1.402\pm0.172$ & $2.716\pm0.176$ & $2.201\pm0.167$ & $3.352\pm0.180$ & $0.197\pm0.005$ & $-42$ & 1\\
$  1.58$ & $1.665\pm0.203$ & $2.755\pm0.200$ & $2.109\pm0.189$ & $3.600\pm0.202$ & $0.211\pm0.006$ & $-42$ & 1\\
$  2.18$ & $2.200\pm0.245$ & $2.843\pm0.249$ & $2.070\pm0.230$ & $3.904\pm0.249$ & $0.212\pm0.007$ & $-42$ & 1\\
$  3.06$ & $1.986\pm0.251$ & $2.830\pm0.256$ & $2.338\pm0.244$ & $3.425\pm0.263$ & $0.207\pm0.008$ & $-42$ & 1\\
$  4.29$ & $2.102\pm0.318$ & $2.532\pm0.355$ & $1.760\pm0.331$ & $3.642\pm0.335$ & $0.216\pm0.010$ & $-42$ & 1\\
$  5.51$ & $2.160\pm0.353$ & $2.734\pm0.387$ & $2.414\pm0.382$ & $3.096\pm0.387$ & $0.209\pm0.011$ & $-42$ & 1\\
$  6.99$ & $2.801\pm0.345$ & $2.960\pm0.381$ & $2.602\pm0.380$ & $3.367\pm0.374$ & $0.210\pm0.011$ & $-42$ & 1\\
$  9.06$ & $1.811\pm0.429$ & $2.483\pm0.502$ & $2.023\pm0.500$ & $3.047\pm0.479$ & $0.168\pm0.014$ & $-42$ & 1\\
$ 12.20$ & $2.361\pm0.670$ & $2.553\pm0.780$ & $2.127\pm0.779$ & $3.066\pm0.751$ & $0.193\pm0.022$ & $-42$ & 1\\
$ -5.25$ & $2.377\pm0.520$ & $2.137\pm0.594$ & $1.212\pm0.502$ & $3.769\pm0.532$ & $0.170\pm0.015$ & $ 48$ & 2\\ 
$ -1.81$ & $2.109\pm0.212$ & $2.654\pm0.205$ & $2.070\pm0.191$ & $3.403\pm0.212$ & $0.199\pm0.006$ & $ 48$ & 2\\
$ -0.73$ & $1.632\pm0.206$ & $2.863\pm0.201$ & $2.003\pm0.182$ & $4.091\pm0.204$ & $0.215\pm0.006$ & $ 48$ & 2\\
$ -0.12$ & $2.300\pm0.172$ & $2.945\pm0.173$ & $2.194\pm0.160$ & $3.955\pm0.175$ & $0.199\pm0.005$ & $ 48$ & 2\\
\noalign{\smallskip}       
\hline
\end{tabular} 
\end{small}
\end{center}     
\end{table*}    

\begin{table*}
\contcaption{}
\begin{center}
\begin{small}
\begin{tabular}{rrrrrrrc}      
\hline 
\noalign{\smallskip}
$  0.48$ & $2.363\pm0.167$ & $2.779\pm0.177$ & $2.102\pm0.166$ & $3.674\pm0.180$ & $0.185\pm0.005$ & $ 48$ & 2\\
$  1.34$ & $1.791\pm0.166$ & $2.885\pm0.188$ & $2.108\pm0.174$ & $3.949\pm0.188$ & $0.175\pm0.005$ & $ 48$ & 2\\
$  4.52$ & $2.083\pm0.346$ & $2.899\pm0.377$ & $2.322\pm0.367$ & $3.619\pm0.371$ & $0.158\pm0.011$ & $ 48$ & 2\\
\noalign{\smallskip}                                                                                        
\hline                                                                                                      
\noalign{\smallskip}                                                                                        
\multicolumn{8}{c}{NGC 67207}\\
\noalign{\smallskip}   
\hline
\noalign{\smallskip}       
$ -7.16$ & $1.855\pm0.440$ & $2.815\pm0.632$ & $2.540\pm0.658$ & $3.119\pm0.592$ & $0.181\pm0.018$ & $-83$ & 1\\
$ -3.87$ & $1.437\pm0.304$ & $2.603\pm0.413$ & $2.127\pm0.417$ & $3.186\pm0.386$ & $0.202\pm0.012$ & $-83$ & 1\\
$ -2.40$ & $2.055\pm0.232$ & $2.552\pm0.328$ & $1.875\pm0.326$ & $3.472\pm0.287$ & $0.212\pm0.009$ & $-83$ & 1\\
$ -1.54$ & $1.542\pm0.192$ & $2.849\pm0.285$ & $1.969\pm0.276$ & $4.121\pm0.247$ & $0.223\pm0.008$ & $-83$ & 1\\
$ -0.93$ & $1.731\pm0.139$ & $3.040\pm0.199$ & $2.178\pm0.197$ & $4.243\pm0.172$ & $0.239\pm0.005$ & $-83$ & 1\\
$ -0.33$ & $1.682\pm0.114$ & $3.097\pm0.164$ & $2.195\pm0.160$ & $4.370\pm0.143$ & $0.249\pm0.005$ & $-83$ & 1\\
$  0.28$ & $1.723\pm0.102$ & $3.118\pm0.149$ & $2.250\pm0.146$ & $4.320\pm0.132$ & $0.246\pm0.004$ & $-83$ & 1\\
$  0.88$ & $1.774\pm0.128$ & $3.187\pm0.186$ & $2.415\pm0.185$ & $4.206\pm0.168$ & $0.227\pm0.005$ & $-83$ & 1\\
$  1.49$ & $1.582\pm0.179$ & $2.706\pm0.273$ & $1.798\pm0.256$ & $4.070\pm0.242$ & $0.213\pm0.008$ & $-83$ & 1\\
$  2.35$ & $1.327\pm0.192$ & $2.548\pm0.304$ & $1.633\pm0.280$ & $3.975\pm0.269$ & $0.200\pm0.008$ & $-83$ & 1\\
$  3.82$ & $2.088\pm0.280$ & $2.491\pm0.415$ & $1.717\pm0.390$ & $3.615\pm0.382$ & $0.192\pm0.012$ & $-83$ & 1\\
$  7.11$ & $1.678\pm0.470$ & $2.746\pm0.683$ & $2.070\pm0.667$ & $3.644\pm0.639$ & $0.198\pm0.020$ & $-83$ & 1\\
$ -5.57$ & $2.197\pm0.608$ & ...             & ...             & ...             & ...             & $  0$ & 3\\
$ -2.25$ & $1.905\pm0.333$ & ...             & ...             & ...             & ...             & $  0$ & 3\\
$ -1.26$ & $1.610\pm0.324$ & ...             & ...             & ...             & ...             & $  0$ & 3\\
$ -0.45$ & $1.814\pm0.324$ & ...             & ...             & ...             & ...             & $  0$ & 3\\
$  0.37$ & $2.149\pm0.313$ & ...             & ...             & ...             & ...             & $  0$ & 3\\
$  1.18$ & $1.669\pm0.315$ & ...             & ...             & ...             & ...             & $  0$ & 3\\
$  1.99$ & $2.124\pm0.365$ & ...             & ...             & ...             & ...             & $  0$ & 3\\
$  5.01$ & $1.833\pm0.468$ & ...             & ...             & ...             & ...             & $  0$ & 3\\
$ -2.81$ & $1.902\pm0.504$ & $2.716\pm0.691$ & $1.857\pm0.662$ & $3.972\pm0.605$ & $0.200\pm0.032$ & $  7$ & 5\\
$ -0.91$ & $1.539\pm0.275$ & $2.746\pm0.322$ & $1.923\pm0.303$ & $3.921\pm0.300$ & $0.234\pm0.016$ & $  7$ & 5\\
$ -0.52$ & $1.719\pm0.124$ & $2.858\pm0.234$ & $2.016\pm0.235$ & $4.054\pm0.192$ & $0.215\pm0.010$ & $  7$ & 5\\
$ -0.24$ & $1.567\pm0.116$ & $2.982\pm0.187$ & $2.141\pm0.192$ & $4.152\pm0.150$ & $0.232\pm0.007$ & $  7$ & 5\\
$  0.04$ & $1.821\pm0.079$ & $3.125\pm0.164$ & $2.292\pm0.155$ & $4.260\pm0.158$ & $0.247\pm0.005$ & $  7$ & 5\\
$  0.32$ & $1.798\pm0.120$ & $3.051\pm0.178$ & $2.220\pm0.181$ & $4.194\pm0.148$ & $0.245\pm0.008$ & $  7$ & 5\\
$  0.60$ & $1.617\pm0.137$ & $2.931\pm0.228$ & $2.121\pm0.226$ & $4.051\pm0.198$ & $0.238\pm0.010$ & $  7$ & 5\\
$  0.88$ & $1.865\pm0.226$ & $2.777\pm0.299$ & $1.980\pm0.310$ & $3.897\pm0.231$ & $0.231\pm0.014$ & $  7$ & 5\\
$  1.27$ & $1.565\pm0.265$ & $2.689\pm0.356$ & $1.943\pm0.366$ & $3.721\pm0.283$ & $0.222\pm0.018$ & $  7$ & 5\\
$  3.19$ & $2.052\pm0.408$ & $2.583\pm0.586$ & $1.741\pm0.564$ & $3.830\pm0.499$ & $0.199\pm0.025$ & $  7$ & 5\\
\noalign{\smallskip}       
\hline
\end{tabular} 
\end{small}
\end{center}     
\begin{minipage}{11.5cm}
{\em Note.} Column (1): radius. Columns (2)-(6): equivalent width of
the line-strength indices. Column (7): slit position
angle measured North through East. Column(8): observing run.
\end{minipage}  
\end{table*}    

\label{lastpage}

\end{document}